\documentclass[sigconf]{acmart}

\AtBeginDocument{%
  \providecommand\BibTeX{{%
    \normalfont B\kern-0.5em{\scshape i\kern-0.25em b}\kern-0.8em\TeX}}}

\copyrightyear{2023} 
\acmYear{2023} 
\setcopyright{rightsretained} 
\acmConference[CHI '23]{Proceedings of the 2023 CHI Conference on Human Factors in Computing Systems}{April 23--28, 2023}{Hamburg, Germany}
\acmBooktitle{Proceedings of the 2023 CHI Conference on Human Factors in Computing Systems (CHI '23), April 23--28, 2023, Hamburg, Germany}\acmDOI{10.1145/3544548.3580999}
\acmISBN{978-1-4503-9421-5/23/04}

\usepackage{calc}  
\usepackage{enumitem}
\usepackage{stfloats}

\usepackage[subtle]{savetrees}

\newcommand{\rev}[1] {{#1}}

\begin{document}

\title[Exploring Challenges and Opportunities to Support Designers in Learning to Co-create]{Exploring Challenges and Opportunities to Support Designers in Learning to Co-create with AI-based Manufacturing Design Tools}

\author{Frederic Gmeiner}
\affiliation{%
    \institution{HCI Institute}
    \institution{Carnegie Mellon University}
    \city{Pittsburgh}
    \state{PA}
    \country{USA}}
\email{gmeiner@cmu.edu}

\author{Humphrey Yang}
\affiliation{%
    \institution{HCI Institute}
    \institution{Carnegie Mellon University}
    \city{Pittsburgh}
    \state{PA}
    \country{USA}}
\email{hanliny@cs.cmu.edu}

\author{Lining Yao}
\affiliation{%
    \institution{HCI Institute}
    \institution{Carnegie Mellon University}
    \city{Pittsburgh}
    \state{PA}
    \country{USA}}
\email{liningy@andrew.cmu.edu}

\author{Kenneth Holstein}
\affiliation{%
    \institution{HCI Institute}
    \institution{Carnegie Mellon University}
    \city{Pittsburgh}
    \state{PA}
    \country{USA}}
\email{kjholste@cs.cmu.edu}

\author{Nikolas Martelaro}
\affiliation{%
    \institution{HCI Institute}
    \institution{Carnegie Mellon University}
    \city{Pittsburgh}
    \state{PA}
    \country{USA}}
\email{nikmart@cmu.edu}

\renewcommand{\shortauthors}{Gmeiner, et al.}

\begin{abstract}
AI-based design tools are proliferating in professional software to assist engineering and industrial designers in complex manufacturing and design tasks. These tools take on more agentic roles than traditional computer-aided design tools and are often portrayed as “co-creators.” Yet, working effectively with such systems requires different skills than \rev{working with complex CAD tools alone}. To date, we know little about how engineering designers learn to work with AI-based design tools. In this study, we observed trained designers as they learned to work with two AI-based tools on a realistic design task. We find that designers face many challenges in learning to effectively co-create with current systems, including challenges in understanding and adjusting AI outputs and in communicating their design goals. Based on our findings, we highlight several design opportunities to better support designer-AI co-creation.
\end{abstract}



\maketitle

\section{Introduction}

\rev{Modern manufacturing processes allow designers to produce complex parts optimizing strength-to-weight or leveraging new materials such as shape-changing plastics, yet their creation often surpasses the designers' cognitive capabilities.}
Recently, computer-aided design (CAD) tools have begun incorporating AI-based \rev{features to generate part designs based on a designer's myriad optimization goals \cite{menges_computational_2011}.}
\rev{For example, Japanese electric vehicle manufacturer WHILL used Autodesk Fusion 360 Generative Design to optimize material economy, strength, and sustainability for an electric wheelchair component \cite{formlabs_generative_2020}. AI tools are also helping designers generate parts using emerging manufacturing processes such as shape-changing smart material structures \cite{yang_simulearn_2020}. In both examples, working with AI allows designers to create designs that would be extremely tedious or infeasible without AI support.}

These generative AI tools take more agency and autonomy in parts of the human-AI design process and are often referred to as “co-creators” \cite{zagalo_enactive_2015}.
\rev{However, effective and practical co-creation presents a significant learning curve for designers, as they are required to work and think collaboratively with AI agents that operate differently than human collaborators or complex CAD tools.}

\rev{Studies have shown that effectively working with professional feature-rich \textit{non-AI} design software already requires substantial and continual learning as such software becomes more capable \cite{mahmud_learning_2020,kiani_i_2020}. 
Consequently, instead of simplifying the software’s interfaces, an active field within HCI studies how the learning of such complex software systems can be better supported through interactive interfaces \cite{lafreniere_these_2015,fernquist_sketch-sketch_2011,masson_supercharging_2022}.}
\rev{However, working with AI "co-creators" is different from working with conventional CAD tools. Designers do not directly manipulate 3D geometry but rather formulate design goals for the AI system to build from. 
Yet, little is known about how to support designers in learning to work with AI tools that take on this more active and collaborative role.}

\rev{In this study, we ask how designers can be better supported in learning to co-create with AI design tools. We guide our investigations 
by a recent wave of HCI literature that looks to inform the design of human-AI collaboration based on the mechanisms that make \textit{human-human} collaborations effective---such as
\textit{grounding in communication} or \textit{shared mental models}  \cite{bansal_beyond_2019, bansal_updates_2019,cai_hello_2019,bittencourt_conceptual_2020,zhou_group_2018,wang_human-human_2020, zhang_ideal_2021}.
Furthermore, we seek inspiration from  \textit{team learning} \cite{wiese_understanding_2019} which models what actions help people \textit{learn} to collaborate effectively with each other.}

\rev{To gain insights on improving human-AI design collaboration, we study how engineering and industrial designers without prior AI co-creation experience learn to work with AI CAD tools in the context of advanced manufacturing design tasks. 
We chose this domain because of its increasing design task complexity, for which designers often \textit{require} AI assistance. 
Through our studies, we aim to generate insights to inform future support interfaces of AI design tools.}
In particular, we \rev{investigate} the following research questions:

\begin{itemize}[font=\bfseries,
  align=left]
\item[\rev{RQ1a}] \rev{\textit{What challenges do designers face when learning to co-create with computational AI tools?}}

\item[\rev{RQ1b}] \rev{\textit{How do designers overcome these challenges?}}
\end{itemize}

To answer these questions, we conducted a series of think-aloud studies observing how trained \rev{engineering and architectural} designers (tried to) learn to co-create with two different computational AI tools on complex manufacturing design tasks (Study 1). Based on analyses of their interactions with the systems and retrospective interviews, we found that they generally valued the AI's assistance but faced challenges in learning to effectively co-create with the tools and interpret the design outputs. Those who were able to produce feasible and satisfying designs learned to co-create with the tool by systematically testing the boundaries of its capabilities early on, by self-explaining AI behaviors they observed, and by sketching and reflecting on design issues.

\rev{After learning about these challenges, we then explored how designers could be supported to better co-create by asking:}

\begin{itemize}[font=\bfseries,
  align=left]
\item[\rev{RQ2}] \rev{\textit{What are effective strategies to support designers in learning to co-create with computational AI tools?}}
\end{itemize}

\rev{To answer this question, we took inspiration from prior work on human-human collaboration \cite{shin_characterizing_2021,meier_rating_2007,burkhardt_approach_2009}. We conducted a human-human collaboration study to see how human guides would assist new users of AI tools in learning to co-create and how the new users learned with human assistance (Study 2).}       
\rev{The observed} effective support strategies included providing step-by-step instructions, prompting design reflection, and suggesting alternative strategies and goals for the design task. We also observed that the human guides relied heavily on multi-modal communication (e.g., screen annotations and mouse gesturing) to communicate more effectively with designers.

\rev{Lastly, to inform design opportunities for new support tools we asked:}

\begin{itemize}[font=\bfseries,
  align=left]
\item[\rev{RQ3}] \rev{\textit{What are designers’ needs and expectations for human-AI co-creation?}}
\end{itemize}

\rev{Synthesizing the results from both studies, we learned that} many participants felt unable to communicate their design goals with the AI and wished for more conversational interactions and contextual awareness from the tool.
\rev{We discuss potential support implications and future work from these needs and expectations.}

In short, this study makes three main contributions: 
\begin{enumerate}
    \item providing a set of observed challenges that engineering \rev{and architectural} designers face when learning to collaborate with AI on complex co-creation tasks in the context of advanced manufacturing design; 
    \item advancing our understanding of designers' needs and expectations for human-AI co-creative tools;
    \item highlighting design opportunities to better support designers in learning to co-create. 
\end{enumerate}

\section{Related Work}

\subsection{AI-based design tools for manufacturing}

AI-based design support tools use various computational methods for generating 2D and 3D design options based on constraints and objectives set by designers \cite{ulu_dms2015-33_2015,allaire_level-set_2002}. In 3D architectural, industrial, and mechanical design, new generative design tools have helped designers create consumer goods \cite{krish_practical_2011}, building layouts \cite{vedaldi_house-gan_2020}, and lightweight automotive and airplane components \cite{noor_ai_2017,de_rycke_nature-based_2018}. In the context of emerging advanced materials, AI design tools assist designers in creating structures out of shape-changing or elasticity-changing materials \cite{yang_simulearn_2020, gongora_designing_2021}.
Such AI tools use multiple techniques to generate designs from a set of goals and requirements, including constraint-based solvers \cite{umetani_guided_2012}, style transfer \cite{atarsaikhan_guided_2020,gatys_image_2016}, simulation and optimization \cite{yang_simulearn_2020}, and genetic algorithms \cite{de_rycke_nature-based_2018}.
Such techniques are becoming commercially available in 3D CAD design tools such as Siemens NX, Solidworks, and Autodesk Fusion360 \cite{siemens_software_generative_2021,dassault_systemes_solidworks_2021,autodesk_fusion_2020}. Many of these tools operate as black boxes where designers first set objectives and then review generated designs.
However, this interaction can make it hard for designers to quickly develop a mental model of how the tool works, limiting their creative use.


Recent research has developed generative design interfaces for interactively exploring multiple design options \cite{zaman_gem-ni_2015,matejka_dream_2018,kim_interaxis_2016} or  more iterative engagement between the designer and the tool through real-time design generation and assessment \cite{kazi_dreamsketch_2017,chen_forte_2018,davis_drawing_2015}.
However, few empirical studies exist that evaluate how engineers and designers \textit{learn} to work with AI design tools on realistic tasks. 
Some existing work has measured the performance impact of AI agents on human engineering teams \cite{zhang_cautionary_2021} while other work has investigated what role professional makers expect for involving AI in digital fabrication workflows \cite{yildirim_digital_2020}. 
This study provides empirical observations on how engineering, industrial and architectural designers learn to work and co-create with computational AI-based design tools.

\subsection{Learning complex software}
Prior HCI research has looked to evaluate and improve the learnability of complex software systems. Past studies explore how people of different ages learn a feature-rich notetaking tool \cite{mahmud_learning_2020}, how professional engineers learn 3D design software \cite{kiani_i_2020}, or how casual designers learn professional motion graphics software \cite{jahanlou_challenges_2021}.
Often, people learn by searching web forums or asking knowledgeable colleagues for help \cite{kiani_i_2020}.
Research on interfaces to support people in learning complex software has proposed dynamic feedforward tool tips \cite{lafreniere_these_2015}, guided tutorial systems \cite{fernquist_sketch-sketch_2011}, and widgets that support self-directed trial and error learning \cite{masson_supercharging_2022}.

\rev{While prior work has explored lenses such as self-directed learning for working with complex software \cite{chaudhury_theres_2022}, it remains an open question of how best to support self-directed learning for co-creation with AI systems that take an active role in the design process.  
For example, prior studies in Human-AI collaboration show that the black box nature of AI systems introduces new challenges where users grapple with non-transparent and non-intuitive system behavior, hindering coordination and communication when completing ``collaborative'' tasks \cite{cai_hello_2019}. To address these issues, various strategies like explainable AI or intelligibility are aimed at helping users refine their mental models of AI systems \cite{miller_explanation_2019,sun_investigating_2022}. 

However, even as AI models become more intuitive for users, we expect that there will always remain a need for learning to work effectively with AI systems to, for example, develop \textit{shared mental models} \cite{kaur_building_2019,zhou_group_2018}. Consequently, as we discuss in the next section, supporting humans in learning to effectively co-create with AI requires bringing in additional theoretical lenses. 
}

\subsection{\rev{Human-human collaboration as a\\lens for studying co-creative systems}}

To design effective human–AI collaboration, researchers have suggested drawing lessons from studying what makes human–human collaboration effective \cite{bansal_beyond_2019, bansal_updates_2019,cai_hello_2019,bittencourt_conceptual_2020,zhou_group_2018,wang_human-human_2020, zhang_ideal_2021,wang_towards_2021}.
While it remains an open question to what extent scaffolds for human-AI collaboration should mirror the designs of supports for human–human collaboration 
\cite{wang_human-human_2020, zhang_ideal_2021}, human-AI interaction researchers suggest that theories and findings from psychology, education, and the learning sciences are currently underutilized. 
For instance, Koch and Oulasvirta \cite{zhou_group_2018} note that \textit{group cognition}---the study of how agents relate to other agents’ decisions, abilities, beliefs, common goals, and understandings---provides powerful concepts for work on human–AI collaboration, yet is rarely referenced within this literature. 
\rev{Group cognition comprises phenomena such as \textit{grounding in communication} \cite{resnick_grounding_1991} (creating mutual sense through verbal and non-verbal communication) and \textit{theory of mind} \cite{engel_reading_2014} (the ability of agents to be aware of their own and the other’s beliefs, intentions, knowledge, or perspectives).}
Similarly, Kaur et al. \cite{kaur_building_2019} argue that \rev{like human-human collaboration,} effective collaborations between humans and AI may require  \textit{shared mental models} between people and the AI to enable \rev{mechanisms such as} adaptive coordination of actions among team members \cite{resnick_grounding_1991,mohammed_team_2001}. 
These may include shared representations of the task to be accomplished, of each other’s abilities and limitations, or of each other’s goals and strategies \cite{van_den_bossche_team_2011, dechurch_measuring_2010, fiore_technology_2016,scheutz_framework_2017}. 
A line of work addressing these opportunities has begun to explore how humans might be supported in developing and maintaining more accurate mental models of an AI collaborator’s capabilities and limitations \cite{bansal_beyond_2019,bansal_updates_2019,kocielnik_will_2019}. \rev{However, compared to concepts of human-human collaboration, honing only users' mental models is not sufficient enough for effective collaboration, which requires \textit{shared mental models} between the user and system \cite{kaur_building_2019}.}

To date, little work has explored how best to support humans in learning to collaborate with AI on authentic tasks, such as design tasks, despite growing recognition of the need for such supports \cite{cai_hello_2019,gero_side-by-side_2019,zhang_ideal_2021,mary_lou_maher_research_2022}. 
Design tasks represent compelling challenges for human–AI collaboration, given that design problems are often ill-defined and require teams to navigate and negotiate both the problem and solution space \cite{dorst_creativity_2001} through an iterative process of generating ideas, building prototypes, and testing \cite{hybs_evolutionary_1992}. 

In this study, we investigate human-AI collaboration for emerging manufacturing design tasks---an area where successful task performance sometimes requires human–AI collaboration, yet where effective collaboration may be challenging to achieve without strong supports \cite{zhang_ideal_2021,gero_side-by-side_2019}.

\subsection{Team learning}

While \rev{phenomena such as \textit{grounding in communication, theory of mind}} and \textit{shared mental models} provide useful concepts to explain which cognitive and social phenomena enable collaboration among a group of agents \cite{zhou_group_2018,kaur_building_2019}, these theories do not explain how groups of individual agents \textit{learn} to effectively collaborate.
To address this gap, \textit{team learning} emerged to study what actions and conditions contribute to how human groups learn to effectively collaborate together \cite{wiese_understanding_2019}. 
For example, team learning studies suggest that the development of effective shared mental models is supported through an \textit{active process of negotiation} between team members, involving \textit{“constructive” forms of conflict}, \textit{argumentation}, and \textit{resolution} \cite{head_effective_2003,jeong_knowledge_2007,van_den_bossche_team_2011}.
However, to date, team learning has been under-utilized as a lens for studying human-AI co-creation. 
In this work, we draw upon concepts from team learning, such as \textit{active processes of communication, joint information processing}, and \textit{coordination of actions} \cite{meier_rating_2007}, to study what actions and support strategies can help designers learn to co-create with AI-based design tools.

\begin{table*}[t]
\caption{Overview of study 1 participants. P-F11* and P-S11* are the same participant who had experience in using both Fusion360 and SimuLearn. Explanation for gaps in participant IDs: Some participants had dropped out after the first design session, or participants were assigned to Study 2 (see Section \ref{sec:guided-sessions}).}
\Description{Study 1 participant demographics. (Table is machine readable)}
\begin{tabular}{llccllcc}
\toprule
\textbf{ID}    & \textbf{Group}     & \begin{tabular}[c]{@{}l@{}}\textbf{Age} \\ \textbf{Years}\end{tabular} & \textbf{Gender} & \textbf{Domain} & \textbf{Occupation}                 & \begin{tabular}[c]{@{}c@{}}\textbf{CAD Exp.} \\ \textbf{Years}\end{tabular} & \begin{tabular}[c]{@{}c@{}}\textbf{Indus. Exp.} \\ \textbf{Years}\end{tabular}\\
\midrule
P-F01  & Fusion360 & 27  & M      & Civil \& Environ. Engin. & Student / MA  & \textgreater 5          & 2 – 5                        \\
P-F02  & Fusion360 & 27  & M      & Mechanical Engineering             & Student / PhD & \textgreater 5          & 2 – 5                        \\
P-F03  & Fusion360 & 25  & M      & Mechanical Engineering             & Student / PhD & \textgreater 5          & 2 – 5                        \\
P-F04  & Fusion360 & 26  & M      & Mechanical Engineering             & Student / MA  & \textgreater 5          & 1 – 2                        \\
P-F05  & Fusion360 & 19  & M      & Architecture, Mathematics          & Student / BA  & 2 – 4                   & 0                            \\
P-F11* & Fusion360 & 64  & M      & Mechanical Engineering             & Contractor    & \textgreater 10         & \textgreater 30              \\
P-F12  & Fusion360 & 59  & M      & Mechanical Engineering             & Designer      & \textgreater 10         & \textgreater 30              \\
P-S01  & SimuLearn & 21  & F      & Architecture                       & Student / BA  & 2 – 4                   & 0                            \\
P-S03  & SimuLearn & 23  & F      & Computational Design               & Student / MA  & 2 – 4                   & 0                            \\
P-S04  & SimuLearn & 21  & M      & Architecture                       & Student / BA  & \textgreater 5          & 1 – 2                        \\
P-S06  & SimuLearn & 23  & F      & Architecture                       & Student / MA  & \textgreater 5          & \textless 1                  \\
P-S07  & SimuLearn & 23  & M      & Architecture                       & Student / MA  & \textgreater 5          & 0                            \\
P-S10  & SimuLearn & 33  & F      & Industrial Design                  & Researcher    & \textgreater 5          & 6 – 10                       \\
P-S11* & SimuLearn & 64  & M      & Mechanical Engineering             & Contractor    & \textgreater 10         & \textgreater 30\\                                                

\bottomrule
\end{tabular}
\label{tab:participants}
\end{table*}

\section{Study 1: Think-Aloud Design Sessions}

We conducted a series of think-aloud studies \cite{van_someren_think_1995} with trained designers \rev{new to working with AI}, where they worked with an AI design tool to complete a realistic advanced manufacturing design challenge. 
\rev{Think-aloud studies have people verbalize their thoughts while performing a task so that researchers can understand their cognitive processes (e.g., forming mental models \cite{ericsson_laboratory_2006}, learning \cite{young_direct_2009}). In our study, we use the think-aloud method to see how AI novices, who encounter a real learning challenge and are less biased than experts, learn to co-create with the AI tools.} 
Participants first completed a 30-minute moderated think-aloud session where a member of the research team observed them working and listened to what they said they were thinking and doing while working.
Half of the participants had a mechanical engineering background and designed a light and strong mounting bracket for a ship engine \rev{while considering the optimal manufacturing method and material combination} using Autodesk's Fusion360 Generative Design (based on topology optimization that generates multiple options) \cite{matejka_dream_2018}.
The other half of the participants with a background in architecture or industrial design designed a bike bottle holder made with shape-changing materials---a complex design task that is challenging to complete without computational support.
Designers working on the bottle holder task worked with SimuLearn \cite{yang_simulearn_2020}, a machine learning-based research tool built on top of Rhino3d that helps designers create structures from shape-changing materials.

After completing the design task, participants submitted their designs and joined a semi-structured interview to reflect on their experience of working with the design tools.
Across the study, we collect the following data:
\begin{itemize}
\item Video and audio recordings and machine-generated transcripts of the open-ended think-aloud design sessions 
\item Audio recordings and machine-generated transcripts of the post-task interviews 
\item 3D designs created during the think-aloud sessions
\end{itemize}

\subsection{Participants}
\label{sec:participants}
We recruited 14 designers (4 female / 10 male, aged 19 to 64 \textit{(M = 32.5, SD = 16.6))} with backgrounds in Architecture, Industrial Design, or Mechanical Engineering (Table \ref{tab:participants}). Most participants were recruited from our institution's student body, but we also recruited three professional designers via an online forum for designers who work with Fusion360 \cite{redditcom_rfusion360_2022}. Participants had a minimum of two years of experience using CAD (Fusion360 or Rhino 3D) but no experience working with the studied AI design tools, determined via a screening questionnaire. Participants included mostly undergraduate and Ph.D. students and three engineers with $>30$ years of industry experience (Table \ref{tab:participants}). 
We recruited participants familiar with either Fusion360 or Rhino3d so that they could focus on learning to work with the AI co-creation features rather than on learning the CAD tool's user interface.
Before the study, all participants signed a consent form approved by our institution’s IRB \textit{(STUDY2021\_00000202)}. Participants were paid 20 USD per hour.

\begin{figure*}[h]
  \centering
  \includegraphics[width=\linewidth]{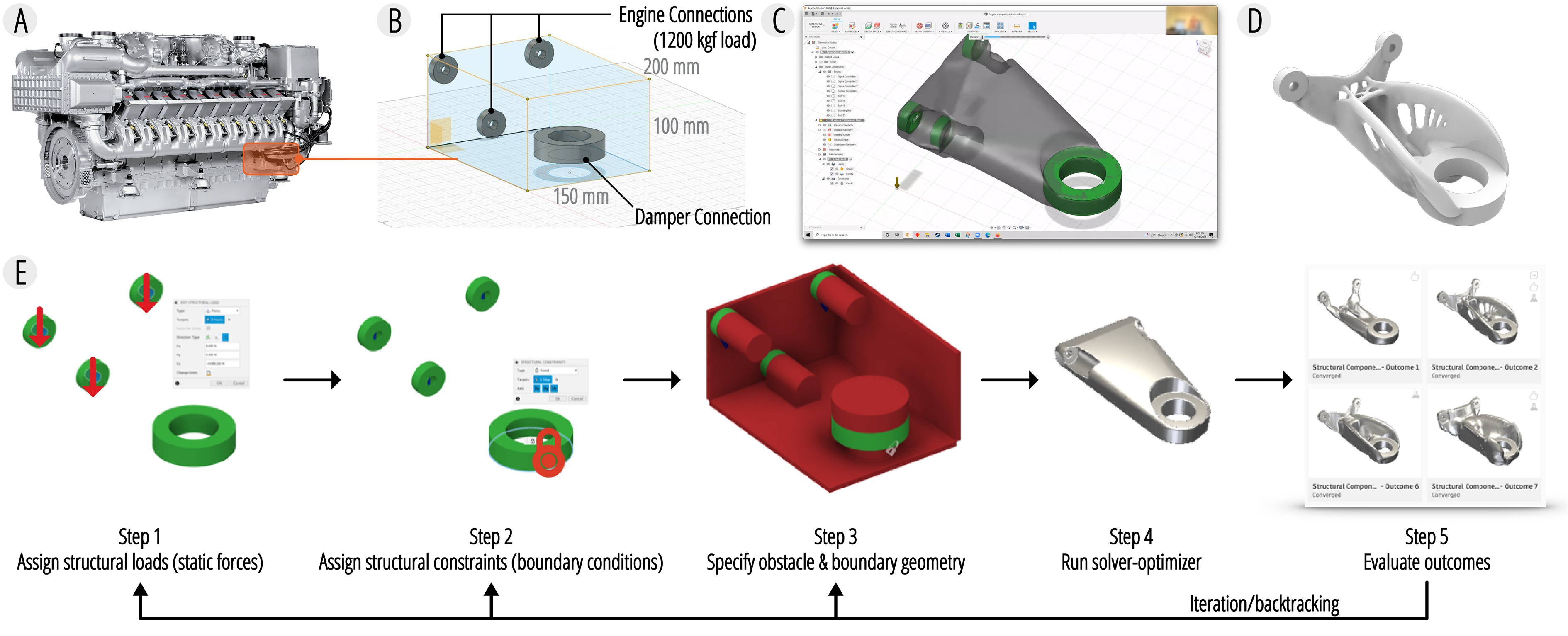}
  \caption{The Fusion 360 design task and workflow. The task involves (A) designing an engine bracket that connects the engine to a damper. (B) A starter file containing connection holes and bounding dimensions is provided to the users to initiate the design in (C) Fusions 360. The user is prompted to create (D) a viable design while minimizing weight. (E) The workflow involves five steps, and based on the AI system's solutions, the user may iterate the design by adjusting the design constraints and criteria to produce new solutions. (Image A: © Rolls-Royce Solutions America Inc.)}
  
  \Description{Process diagram of the Fusion 360 design task steps and workflow steps. Design task steps are illustrated with four images from left to right. The first image shows an engine with a highlighted engine mount bracket connected to the next image on the right. The second image shows a screenshot of a 3D design tool with rings labeled with Engine Connections and Dampener Connections. The third image shows a screenshot of Fusion 360 with a work-in-progress engine bracket. The fourth image shows a rendered image of a generated engine bracket. The workflow steps are illustrated with five images from left to right: The first image shows four 3D rings and a UI form field. The image is labeled “Step 1 Assign structural loads”. The second image shows the same four rings and a different UI form field. The image is labeled “Step 2 Assign structural constraints (boundary conditions).” The third image shows the same four rings together with other enclosing geometry. The image is labeled “Step 3 Specify obstacle geometry.”  The fourth image shows an engine bracket and is labeled “Step 4 Run solver-optimizer.” The fifth image shows a UI with four different engine brackets and is labeled “Step 5 Evaluate outcomes.”}

    \label{fig:fusion-task}
 
\end{figure*}

\subsection{Study context:\\ AI-based design tools and tasks}

To gather generalizable insights into designers' challenges, needs, and expectations around designer-AI co-creation, we observed how designers tried to learn to co-create with \textit{two different computational AI tools} for digital manufacturing tasks. Participants were given a non-trivial, realistic design task to work on during the study. \rev{We selected a mechanical engineering design task concerning multi-dimensional optimization and an industrial design task exploring the use case of shape-changing materials. Both tasks are too complex to accomplish without AI and also have functional AI tools already developed.} We collaborated with engineers and advanced manufacturing experts to identify and pilot the tasks to ensure they were adequately complex but not overwhelming for our target population. The tasks required participants to generate design solutions within a few hours \rev{over multiple sessions.}

\subsubsection{\textbf{Mechanical design support tool}}
Mechanical designers worked with the "Generative Design" feature of Autodesk Fusion360\cite{autodesk_fusion_2020}, which helps designers to create lightweight and strong parts through topology optimization and genetic algorithms. In this task (Figure \ref{fig:fusion-task}A-D), the designer is asked to design a material-efficient and structurally-sound engine mounting bracket \rev{by considering the optimal manufacturing and material combination from a large pool of possibilities. While designing mounting brackets is common for mechanical engineers, optimizing designs for different manufacturing methods and materials is difficult without simulation and AI support. Traditionally, engineers would first build a part and then gradually remove or add material based on structural analysis to derive a weight-optimized part. Exploring different manufacturing options would be necessary for every material and manufacturing constellation---which is time-consuming and tedious. In contrast, Generative Design can automatically generate many different design options based on specified high-level requirements, which the designer can explore and choose from. }

Participants were provided a starter file containing the geometric constraints and needed to specify the mechanical design criteria (e.g., loads, bolt connection clearance, boundary condition). 
Participants then ran the solver and evaluated the AI-generated solutions to identify three viable designs for submission (Figure \ref{fig:fusion-task}E). If none of the outcomes were deemed satisfactory, the user might choose to iterate the design by adjusting the input criteria.  

\subsubsection{\textbf{Industrial design support tool}}
Industrial and architectural designers worked with SimuLearn \cite{yang_simulearn_2020}, a research system built on Rhino3D that uses ML-driven simulation and optimization to \rev{enable} designers to rapidly create objects out of shape-changing materials. This manufacturing process creates 3D-printed flat grids out of PLA plastic that can transform into a volumetric shape when heated. The transformation and the resulting shape can be controlled by tuning the grid geometry and the portion of the active transformation element (i.e., actuator ratio) within the beams. 
This technique is envisioned to reduce production waste and shipping costs (e.g., flat packaging, reduced support material). 

\rev{While new materials are being developed to manufacture shape-changing structures, designing shape-changing components poses unique challenges that designers are ill-equipped to handle. Unlike 2D and 3D design, designing with shape-changing materials involves a non-intuitive mismatch between the final target (3D shapes) and the design input (often 2D).
Effectively designing for shape-changing materials requires an understanding of (often complex) spatiotemporal, self-assembling material behaviors that may push against the limits of what humans can mentally simulate. At the low level, designing such materials requires modifying volumes voxel-by-voxel, which is infeasible for complex structures if done manually. AI-driven tools allow designers to create complex artifacts that would otherwise be impossible to create by hand.}
SimuLearn aids the design process by providing real-time simulation and optimization to iterate designs toward the desired morphing behavior.

\begin{figure*}[h]
  \centering
  \includegraphics[width=\linewidth]{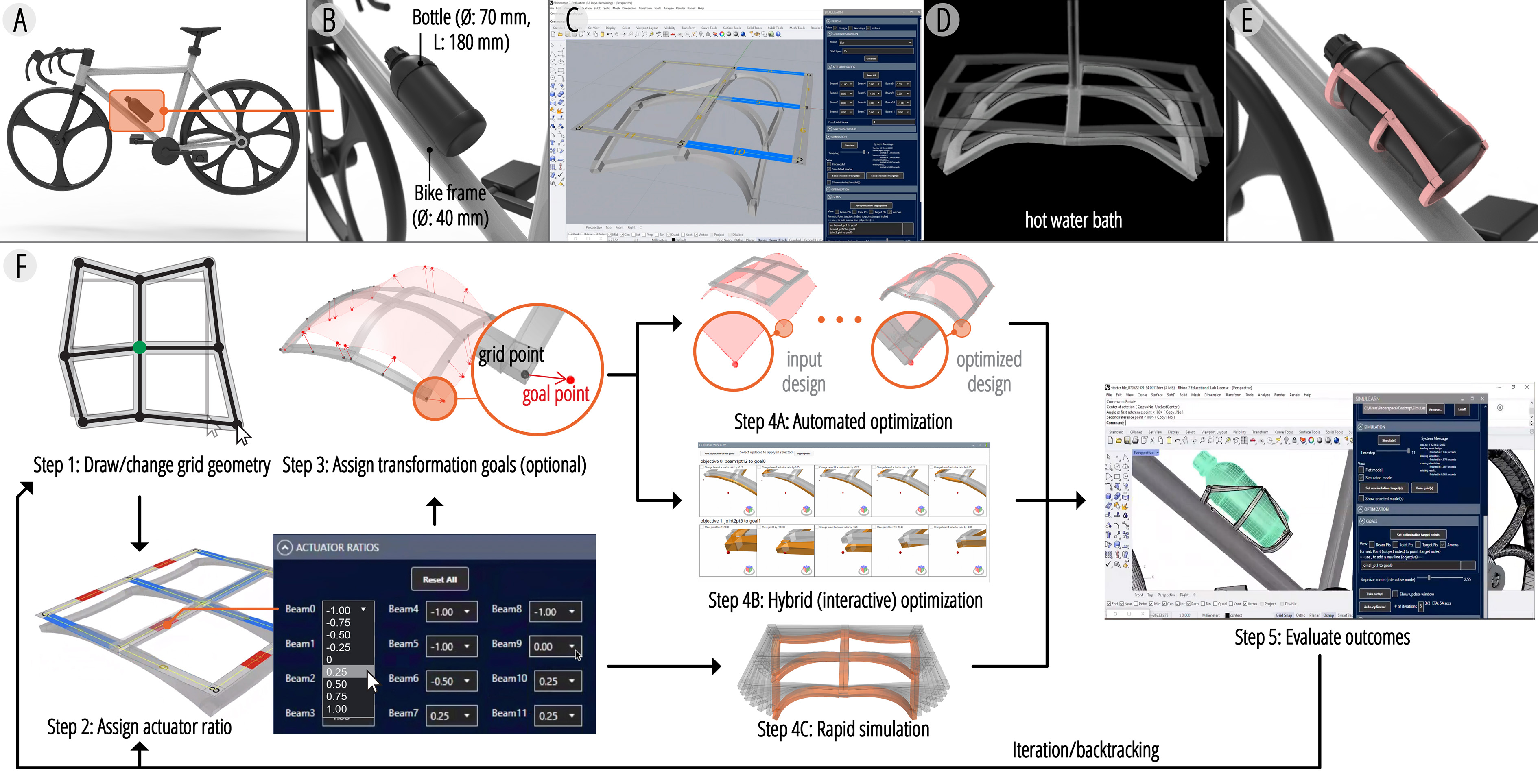}
  \caption{ The SimuLearn design task. (A) The user is prompted to design a bike bottle holder using the provided (B) starter file and (C) the SimuLearn tool to create (D) morphing grid structures. (E) The transformed grid should assemble into the holder and fit around the bottle and bike frame. (F) The SimuLearn tool provides three types of design workflows. After initializing a morphing grid design, the user can choose between AI-assisted optimization functions or rapid simulations to iterate the design.}
  \Description{Process diagram of the SimuLearn design task steps and workflow steps.
Design task steps are illustrated with five images from left to right. The first image shows a 3D rendering of a bike with a water bottle next to the frame, which is connected to the next image on the right. The second image shows a close-up of the water bottle. The third image shows a screenshot of Rhino3d and the SimuLearn interface. The fourth image shows a deforming grid made of shape-changing material labeled with “Hot water bath.” The workflow steps are illustrated with seven connected images: The first image shows a 2x2 grid and a mouse cursor that moves the lower right corner of the grid to the right. It is labeled with “Step 1: Draw/change grid geometry.” The second image is labeled “Step 2 Assign actuator ratio” and shows the SimuLearn UI next to the 2x2 grid. The third image is labeled “Step 3 Assign transformation goals” and shows a bending 3D grid structure. The fourth image is labeled “Step 4A Automated Optimization” and shows two deformed grids side by side. The fifth image is labeled “Step 4B Hybrid interactive optimization” and shows a UI window with ten viewports containing different grid deformations. The sixth image is labeled “Step 4C Rapid Simulation” and shows superimposed different states of a morphing grid. The seventh image is labeled “Step 5 Evaluate outcomes” and shows a screenshot of Rhino3d and the SimuLearn UI.}
  \label{fig:simulearn_task}
\end{figure*}

In this task (Figure \ref{fig:simulearn_task}A-E), participants design a bike bottle holder using morphing grids. A starter file containing the bottle and bike frame geometry was provided to contextualize the design. This task was more open-ended than the mechanical engineering task since each designer may assemble the morphed grids in different ways to create the holder. To use the tool (Figure \ref{fig:simulearn_task}F), the designer models the grid geometry and assigns bending actuator ratios. Next, participants simulate the design, observe the predicted transformation, and iterate the design by changing the grid model and actuator assignment. Alternatively, participants may opt to use functions to optimize the grids toward a targeted transformed shape. The optimization process can be either autonomous or interactive (i.e., the tool suggests edits for the user to choose from). \rev{To effectively work with the tool}, users need to learn to work with the different levels of AI assistance to produce a satisfactory design iteratively.

\subsection{Choosing design tools}
We specifically study these systems for two reasons. First, while both tools support advanced manufacturing tasks, they represent computational systems with distinct purposes and interaction paradigms. 
Fusion360's Generative Design module assists engineering designers with the generally familiar task of creating light and structural solid parts. The AI system helps designers to navigate a large design space and explore opportunities while adhering to specified requirements and constraints.
SimuLearn, on the other hand, supports designers in working with an emerging material and manufacturing process unfamiliar to most designers. SimuLearn's AI tool provides rapid simulations of the shape-changing material and offers different levels of design assistance---from manual, over interactive, to autonomous optimization/iteration.

Second, each tool represents a different interaction style and synchronicity. In Fusion360, users follow a structured sequence of steps to set up the parameters and acquire solutions. It may take a few hours to generate new solutions, and the designer may export the generated models at any time or iterate the design by adjusting the parameters and rerunning the solver. By contrast, SimuLearn's solver runs two to three magnitudes faster (5-180 seconds), and users interact with the system without a predefined workflow. Participants may also freely switch between the three levels of AI support at any point.

\newpage

\section{Study design}

\begin{figure}[b]
  \centering
  \includegraphics[width=\linewidth]{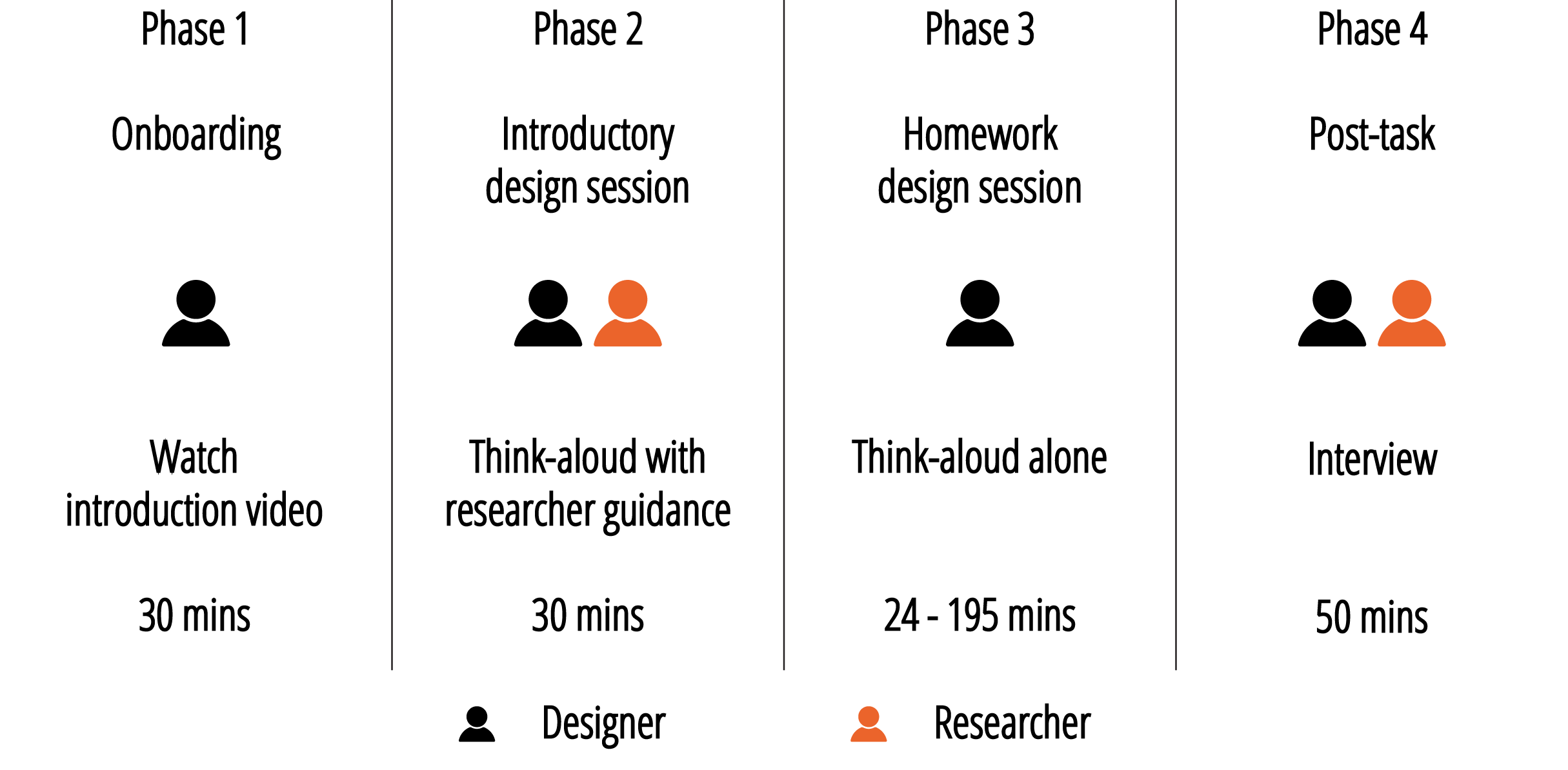}
  \caption{Overview of Study 1 think-aloud design sessions procedure. Participants were first introduced to the design tool and task, then worked while thinking aloud across multiple sessions. They completed an interview after turning in their design.}
  \Description{Schematic illustration of process steps of the study procedure. Four columns in total are labeled Phase 1, Phase 2, Phase 3, and Phase 4. Phase 1 has labels Onboarding, Watch introduction video, 30 minutes. Phase 2 has labels Introductory design session, Think-aloud researcher guidance, and 30 minutes. Phase 3 has labels Homework design session, Think-aloud alone, 24 to 195 minutes. Phase 4 has labels Post-task, Interview, 50 minutes.}
  \label{fig:study_procedure}
\end{figure}

\subsection{Study procedure}

The study was structured into four phases (see Figure \ref{fig:study_procedure}):

\textbf{1) On-Boarding:} Before the first session, participants received instructions by email on how to access the design tools running on a dedicated remote machine. They also watched a 25-minute video tutorial demonstrating the tools' core functionalities with a step-by-step example.

\textbf{2) Intro Design Session:} At the beginning of the first session, a research team member introduced the participants to the study, design brief, task, and starter file. Sessions were conducted over video conference (Zoom) with audio and video recording. Participants then worked while sharing their screens and thinking aloud. \rev{Participants were allowed to use all available support resources, such as internal help files, external video tutorials, or user forums.} The researcher quietly observed the participants setting the parameters of the computational design system and took notes. 
\rev{Due to the complex and open-ended nature of the tasks, the research team interrupted the task after 30 minutes and asked participants to continue working on their own as a compensated "homework assignment" without any time limit. }

\textbf{3) Homework Sessions: } Following the initial session, participants continued working independently for as long as needed to iterate and produce the final design submission. Participants used the same remote machine and joined a personalized video conference meeting with automatic recording to document their work while thinking aloud. We also deployed a simple web application to prompt the users to verbalize their thoughts. The application analyzed the speech input from the microphone and reminded the user to "keep talking" after twenty seconds of silence. Once the designer felt satisfied with the result, they submitted the 3D files to our team.

\textbf{4) Post-Task Interview:} Within two days of submitting their result, participants completed a one-hour semi-structured interview with a research team member. Participants were asked to reflect on their experience working with the AI tool. The interviews were conducted remotely over video conferencing. The interviewer took notes, and the interview audio and video were recorded. The interview protocol contained 36 questions clustered into three topics: \textit{collaboration with the tool}, \textit{design process}, and \textit{learning process} (see the Appendix for interview protocol). \rev{The topics and questions clustered under \textit{collaboration with the tool} were inspired by measures from team learning literature on assessing collaboration quality and effectiveness of human teams \cite{van_den_bossche_team_2011,jeong_knowledge_2007,meier_rating_2007,burkhardt_approach_2009}. These include perceived team roles and coordination, communication between user and tool, conflict resolution, timing, and (shared) mental models. }  

\subsection{Analysis}
\rev{To gain insight into research questions 
\textbf{RQ1a} \textit{What challenges do designers face when learning to co-create with computational AI tools?} and  
\textbf{RQ1b} \textit{How do designers overcome these challenges?} we (1) evaluated the design outcomes and analyzed more than 40 hours of think-aloud videos and 17 hours of interview recordings using
(2) video interaction analysis of think-aloud videos, and (3) reflexive thematic analysis of think-aloud sessions (videos, transcripts) and interview transcripts. }

\subsubsection{\textbf{Evaluation of design outcomes}}
We evaluated the effectiveness of Designer-AI collaboration by measuring the time required to complete the task and designer satisfaction with their results as rated on a three-point Likert scale (satisfied, neutral, unsatisfied) during the post-task interview. We also measured product feasibility for the mechanical engineering task by checking the designed engine brackets against the requirements in the design brief. The structural soundness was validated using finite element analysis (FEA), and the used material was checked by measuring part volume. We also checked the models for shape requirements (i.e., clear bolt holes, body within the bounding box). Since the bottle holder was a more free-form and aesthetic design task, we only checked if the user submitted their design and primarily relied on their self-reported satisfaction with the outcome.

\subsubsection{\textbf{Video interaction analysis}}
We used \textit{video interaction analysis} \cite{baumer_comparing_2011} of the think-aloud recordings to understand participants' learning process while working with the AI features. To understand how well participants learned over time to use the AI features effectively, we tracked their interactions with the AI features relevant to the design task and documented whether the actions would produce satisfactory outcomes. 
For Fusion360, we tracked how participants specified structural loads, mechanical constraints, and the obstacle geometry feature to control the bracket's bolt clearance and overall size. For SimuLearn, we tracked how participants used different AI-assisted features (hybrid and automated optimization)  throughout the think-aloud sessions. 

\subsubsection{\textbf{Reflexive thematic analysis}}

To understand participant's challenges, needs, and expectations when learning to co-create with the AI system, we performed a \textit{reflexive thematic analysis} \cite{braun_reflecting_2019} of the interview data (transcripts) and the think-aloud sessions (video, transcripts). We followed an iterative inductive coding process and generated themes through affinity diagramming. We used ATLAS.ti to analyze transcripts, audio, and video. 

In the initial coding, the think-aloud and interview transcript data were equally distributed among two researchers who generated preliminary codes utilizing both a \textit{semantic} (what people said) and \textit{latent} (our interpretations of the data) coding strategy. Next, the research team collectively identified initial codes and themes.
\rev{We generated themes in a bottom-up manner. However, we looked at the data with a mindset of collaboration between the designer and the tool---inspired by previous studies on human-human collaboration, co-creation, and team learning \cite{meier_rating_2007,burkhardt_approach_2009}. We also tried our best to identify and separate usability issues from the codes and themes to avoid confoundment.}

The two researchers then coded the think-aloud recordings to document where designers specified system parameters or evaluated design outcomes. These moments allowed us to find many of the problems that designers faced. We also coded non-verbal expressions like mouse gesturing or screen annotations that showed how designers attempted to communicate.    

Finally, we created \textit{summary videos} highlighting specific situations \rev{related to co-creation with the tool} (e.g., designers confused by AI-generated outcomes). The video clips were annotated with a time code, participant ID, and a contextual description of the situation to share and discuss with the entire research team (for an example, please see the video figure in the supplementary material). 
The research team collectively analyzed the think-aloud summary videos in a half-day session and discussed the themes. We completed the qualitative analysis by iteratively reviewing and revising codes and themes until we identified a stable network of coherent and rich themes.

\begin{table}[h]
  \centering
    \caption{ Evaluation of design outcomes for the engine bracket design task (left) and bottle holder design task (right). The designer's satisfaction with the outcome is rated with green=satisfied, yellow=neutral, red=unsatisfied. For the engine bracket task, meeting structural and shape requirement checks are rated as X=fail, check mark=pass.}
  \Description{Two tables showing the summative data of the design sessions.}
  \label{tab:design_evaluation}
  \includegraphics[width=\linewidth]{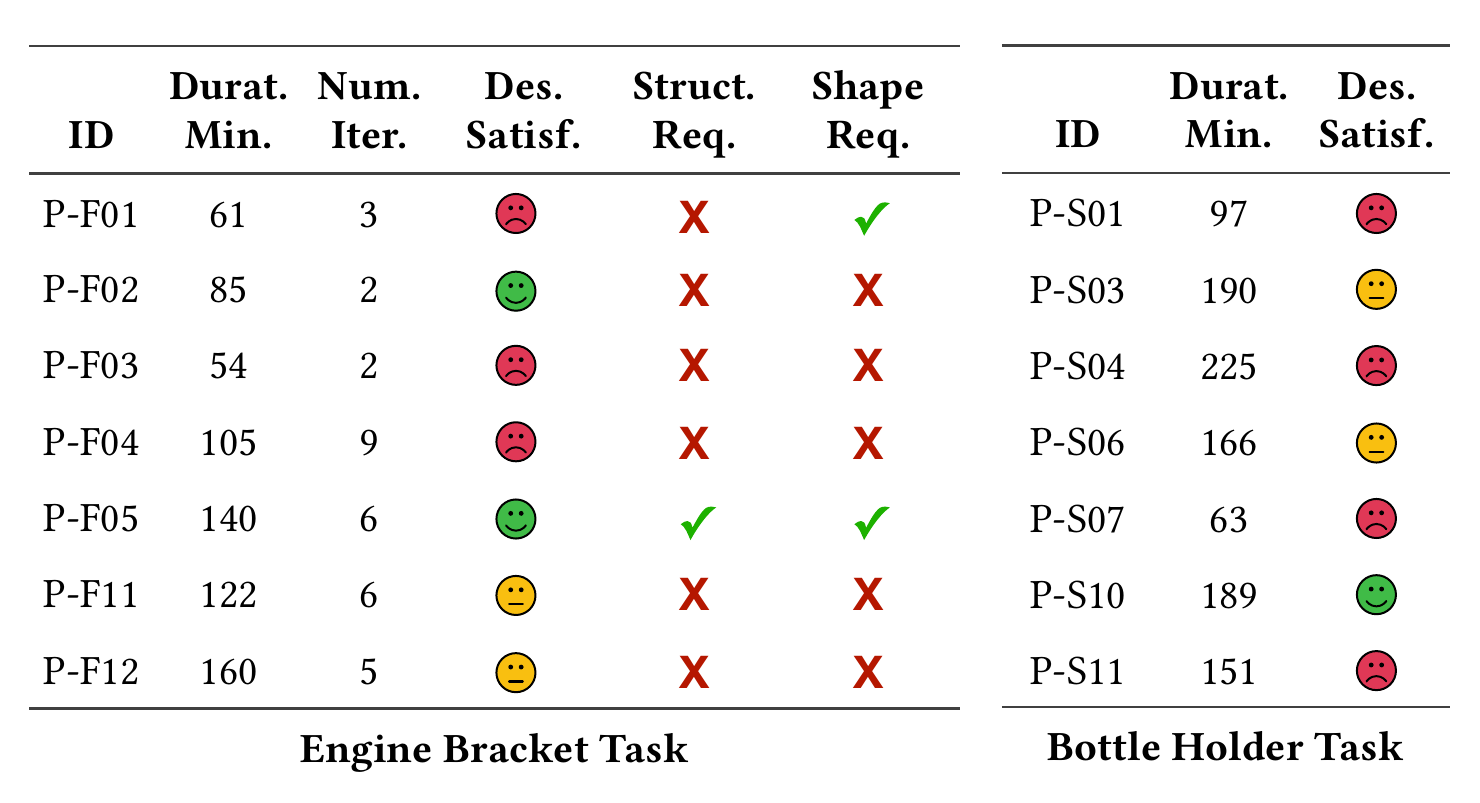}

\end{table}

\begin{figure}[h]
  \centering
  \includegraphics[width=\linewidth]{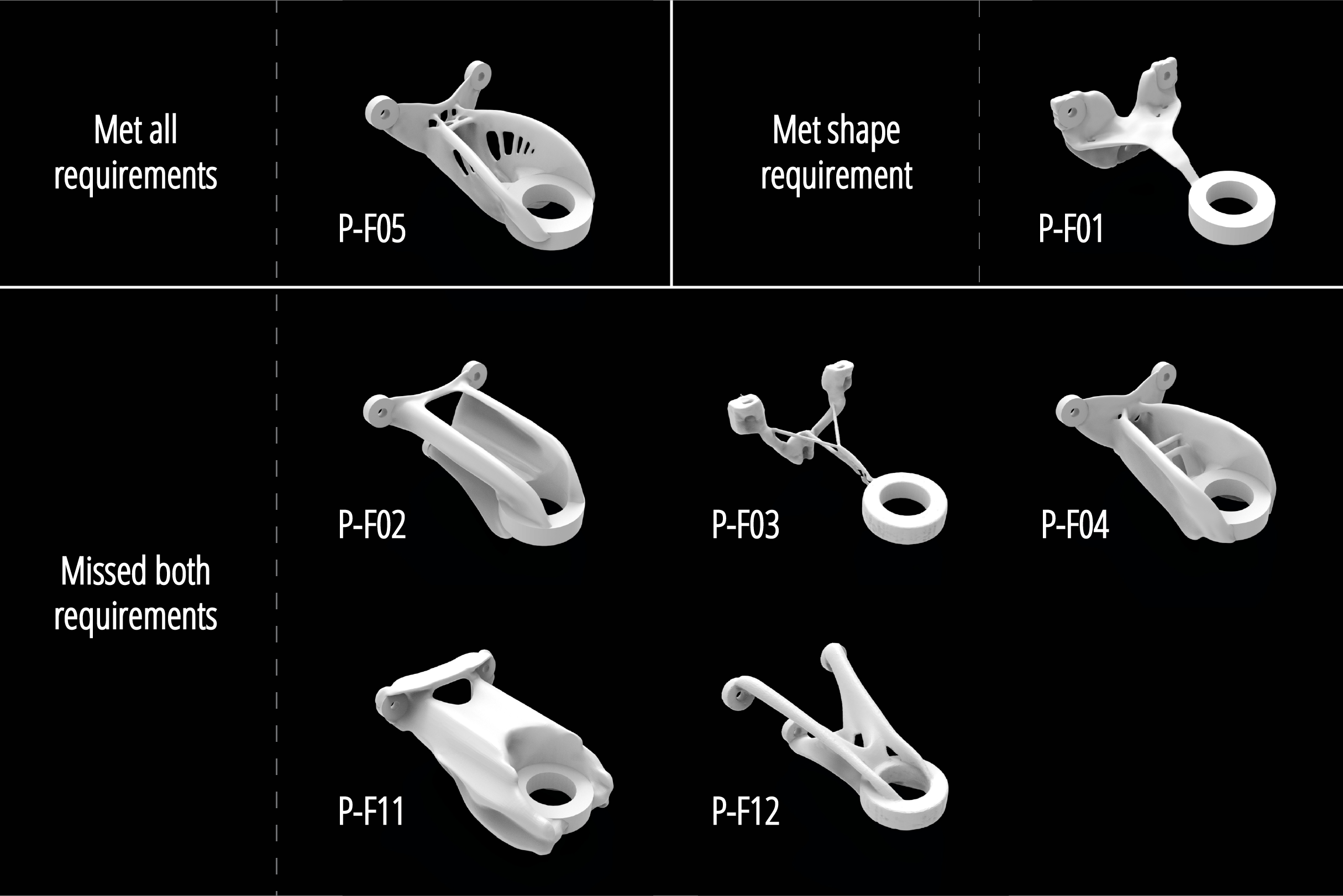}
  \caption{Overview of engine bracket designs created by participants. P-F05 met both the shape and structural requirements. P-F01 only met the shape requirements (the bracket has holes for bolts with sufficient clearance and respects the specified dimensions but is structurally too weak). All other brackets missed both requirements and were either too heavy, weak, larger than the specified dimensions, or had not enough bolt clearance. For simplicity, we only show one design option per participant. Please see the Appendix for all submitted design options.}
  \Description{Renderings of seven engine bracket designs created by participants. The image is split into three areas. The first area is labeled “Met all requirements” and contains one engine bracket design. The second area is labeled “Met shape requirements” and contains one engine bracket design. The third area is labeled “Missed both requirements” and contains five engine bracket designs.}
  \label{fig:brackets_designs}
\end{figure}

\begin{figure}[h]
  \centering
  \includegraphics[width=\linewidth]{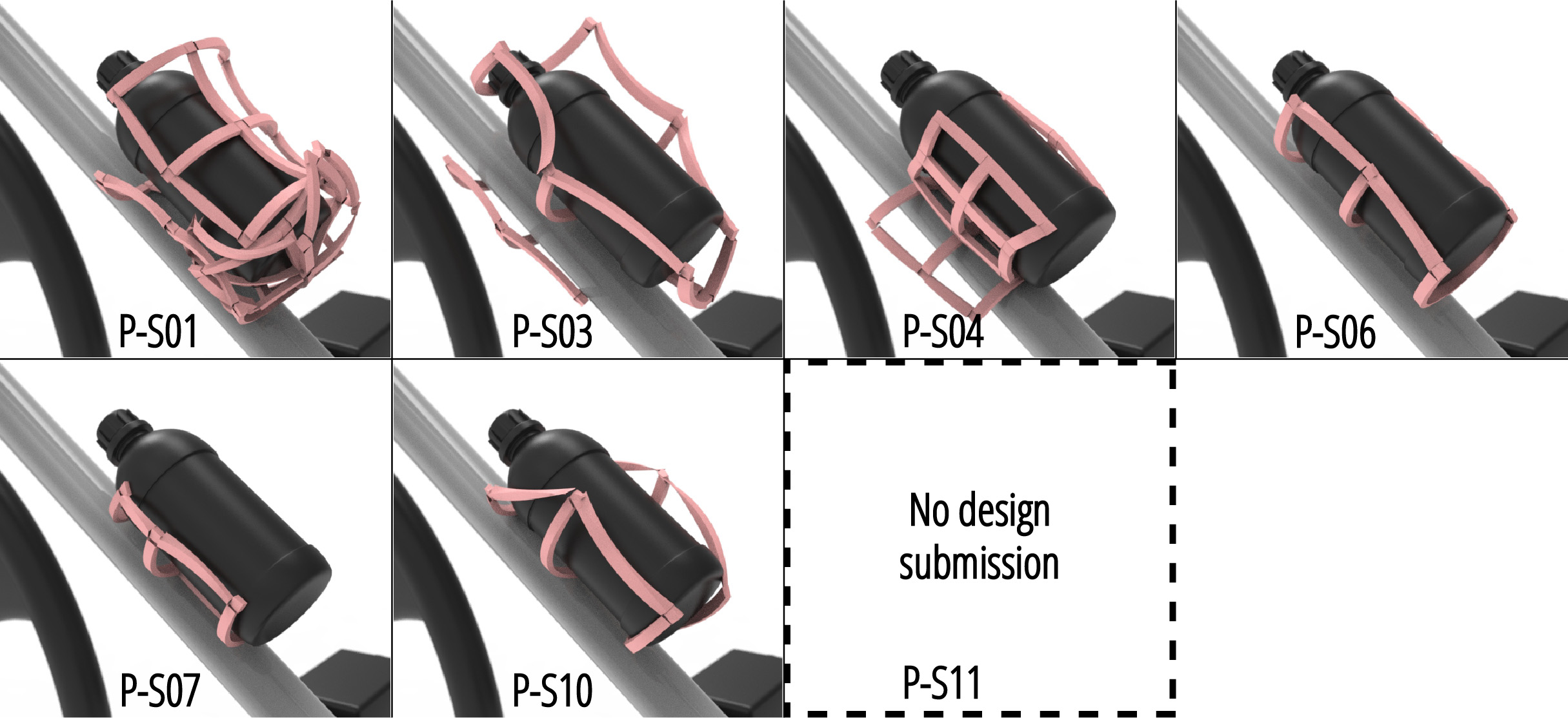}
  \caption{Overview of bottle holder designs created by participants. Participants came up with different designs using the shape-changing grid material in various ways. Participant P-S01 was not able to control the shape as intended and improvised by stacking grids together. While some designs (P-S06, P-S07) would not be able to hold a bottle, some designers (P-S04, P-S10) utilized the shape-changing grid structure to create promising bottle holder designs. Participant P-S11 stopped working on the task out of frustration and did not submit a design.}
  \Description{Renderings of six bike bottle holders created by participants.}
  \label{fig:bottle_holder_designs}
\end{figure}

\section{Results}

\rev{Overall, participants expressed seeing potential value in the AI systems to support their design process---especially that the tools would enable them (at least in theory) to create and explore more complex designs in a shorter time than without AI.} However, most faced unresolvable challenges in learning to effectively co-create with the tools. In the following subsections, we provide a brief overview of participants' performance on the design tasks and then explore what challenges they faced in co-creating with them \rev{(RQ1a)}. We then highlight what some successful learning strategies looked like \rev{(RQ1b)}. 

\subsection{Design Tasks Outcome Summary}

\rev{In the engine bracket task, all participants were familiar with designing similar mechanical components by considering forces and constraints. Generally, such a task is a standard exercise in engineering education, and our task was comparable to the example provided by Autodesk in the introduction video that the participants watched.}
Participants required between 54 and 160 minutes \textit{(M=104, SD=39.7)} to complete the task (see Table \ref{tab:design_evaluation} left).
\rev{No participant mentioned in the post-interview that the task itself was too difficult for them.} Yet, only participant P-F05 was able to produce a self-satisfactory design that met both shape and structural requirements (see Figure \ref{fig:brackets_designs} and Appendix for additional designs).
P-F02 was also satisfied with their design but opted to manually refine the generated geometry that did not meet the requirements (i.e., using excessive materials and blocked bolt holes).
\rev{We were surprised to find that few engineering participants produced satisfying results, even though they were familiar with the type of design task. Designers struggled to perform this otherwise familiar design task when they attempted to do so with AI assistance.}

\rev{For our industrial or architectural designers, designing a bottle holder in 3D was not perceived as difficult. However, working with shape-changing material structures was new and everyone expressed in the post-interview that working with the shape-changing material was "unintuitive" and "challenging." } 
Participants worked on the task between 63 and 255 minutes in total \textit{(M=154, SD=56.6)}. All but one participant submitted a bottle holder design (Figure \ref{fig:bottle_holder_designs}). This participant stopped working on the project after 151 minutes because he felt he could not control the material well, even with the AI. In the end, almost all designers were either dissatisfied with their final design or had a neutral opinion (see Table \ref{tab:design_evaluation}).

\begin{table*}[h]
  \centering
    \caption{Schematic overview of each participant's learning process of design task-relevant features of the AI system. Left: Engine Bracket Task. Right: Bottle Holder Task. Correct use of input parameters or AI tools is shown with check marks. Increasing numbers of check marks from the first iteration to the last iteration suggest participants learning to work with the AI system.}
  \label{tab:learning_process}
  \includegraphics[width=\linewidth]{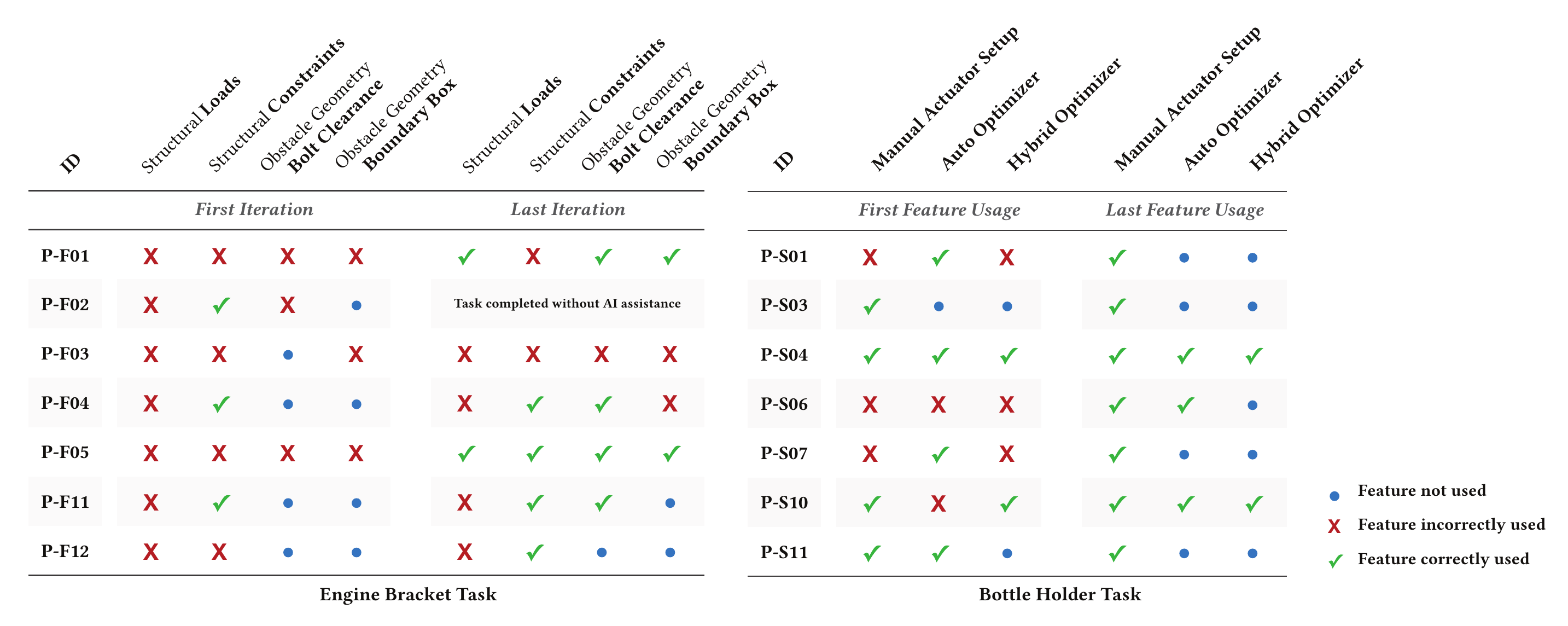}
  \Description{Two tables with a schematic overview of each participant’s learning process of design task-relevant features of Fusion360 (left) and SimuLearn (right).}
\end{table*}

Table \ref{tab:learning_process} shows that only a few designers understood how to set relevant parameters for the design task to produce satisfying results on the first attempt in both tasks. 
In Fusion360, most designers learned over time to successfully specify the structural constraints and obstacle geometry for bolt clearances. However, many designers failed to correctly specify structural loads and the boundary box. As a result, some submitted designs were too heavy, weak, or larger than the specified dimensions. In particular, one designer (P-F02) decided to manually refine the design after the first iteration because they felt more in control this way. For SimuLearn, all participants quickly learned how to control the grid shape using manual adjustments and rapid simulations (Table \ref{tab:learning_process} right). However, many avoided using the other two AI-assisted features after an unsuccessful first attempt and continued to work manually (see Section 5.2.2 for further explanation). 

\subsection{Challenges designers faced when learning\\ to co-create with AI \rev{(RQ1a)}}

We clustered challenges designers faced when learning to co-create into three themes: \textit{Understanding and fixing AI outputs}, \textit{working ``collaboratively'' with the tools}, and \textit{communicating their design goals to the AI}. 

\subsubsection{\textbf{Challenges in understanding and adjusting AI outputs}}
Designers were often confused about the generated results and had difficulty understanding the cause and remedy of ``weird'' outcomes. This often occurred when generated solutions contained minor aesthetic flaws such as surface bumps, holes, or slightly twisted geometry. Designers wondered about the AI's underlying design rationale and were unable to determine whether such design features were intended or caused by algorithmic glitches. 

Furthermore, designers were unsure how to correct the generated designs (e.g., identifying the parameters that led to the problem). Some designers were hesitant to manually refine the generated geometry because they felt uncomfortable changing the optimized structure:
\begin{quote}
\textit{"...I also realized that if I was making any change in the mesh maybe I was changing the stress that that area will have. So I didn't feel very comfortable changing stuff because I knew that was optimized for my loads and my material."} P-F04
\end{quote}
Designers were also unable to make sense of apparent structural issues in the generated designs, like when the generated parts were unreasonably thin or thick, provided little clearance, or produced confusing grid transformations. As verbalized by this designer while evaluating a generated engine bracket:

\begin{quote}
    \textit{“it just doesn't make sense that, like this region here, which is obviously pretty thick... well I guess it's not. I just don't think that it can hold up the weight. It says the factor of safety is five. That just seems ridiculous to me.”} P-F01 Think-aloud
\end{quote}

At other moments, designers were uncertain whether the AI or the user was responsible for fixing the problem. In some cases, designers accepted imperfect results and attributed the flaws to the system. On one occasion, a designer verbalized their concerns when evaluating a generated engine bracket with insufficient bolt hole clearance:

\begin{quote}
    \textit{"It's a goofy-looking bracket. I would never design it like that, but this thing thinks it can do it like that […] It just seems [that there needs] to be bolt heads and stuff, so I just don't see that bolt fitting in this area. But this is what it gave me, so I don't have a problem with that.”} P-F13 Think-aloud
\end{quote}

\subsubsection{\textbf{Challenges in working ``collaboratively''}}

Designers frequently expressed challenges in sharing control with AI-based design tools, noting that it often felt like these tools were dominating the design process. In those situations, designers either gave up and accepted unsatisfying results, improvised 'hacky' strategies to work around the AI or abandoned the AI assistance altogether and proceeded to work manually.
For instance, one designer using SimuLearn expressed frustration about having insufficient control over the design process and decided to accept imperfect results:
\begin{quote}
    "\textit{I feel like the collaborative process [...] it seemed a little difficult to control. I felt that SimuLearn had more control over it than I did.}" P-S01 Interview
\end{quote}
Similarly, another participant pointed out the lack of participation opportunities and conversation throughout the design process:
\begin{quote}
    \textit{"I would say no, that it's not co-creative. The whole program creates the thing but based on my limits. [...] I don't feel like I interact in that creation of the shape. I just worked at the beginning and then I let the program do the rest. I missed that once there is one option, I cannot change anything, I cannot interact with the solutions that the program is giving me."} P-F04 Interview
\end{quote}

In other cases, designers developed their own 'hacky' workarounds to attain feasible results. 
In Fusion360, this happened when designers tried to use simple boxes as obstacle geometries to limit the material growth within the specified perimeter. However, the algorithm often found ways to ``squeeze'' in additional, unwanted material through the gaps between obstacle geometries. A Fusion360 user expressed they felt like tricking the system when they were unsuccessful at blocking material generation:
\begin{quote}
    \textit{“I think the software did kind of dominate the design process a little bit in that I was making things to satisfy the software instead of it kind of adapting to my needs [...] I think for something that's like supposed to be so smart and easy to use I wish it gave me more options instead of [me] trying to find these like little tricks. So I don't want to use little tricks. I don't want to use a hidden kind of ‘I'll trick the software into thinking that it's correct.’"} P-F05 Interview
\end{quote}

Finally, designers often avoided AI assistance when they could not learn to co-create with it effectively. This issue was particularly frequent when designers used SimuLearn, given that this tool explicitly allows designers to switch between different levels of AI assistance. When comparing which SimuLearn features designers used to complete the tasks, all designers learned over time to manually adjust, simulate, and iterate the design (see Table \ref{tab:learning_process} right). However, only two designers (S04, S10) were able to work with the AI-assistance features (i.e., auto and hybrid optimizer) on the first try. All others expressed that they could not make sense of the AI output, even when they used the feature correctly. These participants subsequently avoided using the AI features and continued to work manually with rapid simulation.

\subsubsection{\textbf{Challenges in communicating design goals to the AI}}

Designers often had difficulty communicating design goals to the AI system. For example, designers were unsure about the use and implication of certain parameters. Furthermore, many designers recognized their knowledge gaps of parameters that defined manufacturing processes or materials. E.g., a designer was unfamiliar with the specificity of a manufacturing method and kept using default parameter values:

\begin{quote}
    \textit{"I'll have to see if they have five-axis milling... minimum tool diameter 10 millimeters... uh I don't know, I'm not really a tooling guy... tool shoulder length... [laughs] minimum tool diameter... all right I guess I don't know... I just hit okay on that."} P-F12 Think-aloud
\end{quote}

Designers often relied on the AI system's default settings or made assumptions about their effect when they were unsure about the parameters' meanings. Instead of seeking clarification from helpful resources, they often tried to determine a parameter's effect on the final result independently. However, it was oftentimes hard or impossible to notice and trace back parameter influences from the final results.

In the think-aloud sessions, we also observed several occasions where designers had different interpretations of the parameters. For example, in the Fusion360 task, half of the users made mistakes when converting the loads into the correct unit expected by the system. Similarly, when applying loads to multiple targets (i.e., bolt holes), the system applied the same load to each of the targets instead of equally distributing the load across the targets, which the designers had anticipated \footnote{This is a known issue in Fusion360 that many users have discussed in the user forum. See \cite{autodesk_generative_2021}.}. This mismatch led to higher load assignments and unnecessarily strong and bulky bracket designs. Interestingly, most designers verbalized their uncertainty about the load distribution when specifying parameters, as exemplified by this think-aloud comment: 

\begin{quote}
    \textit{“All the loads... I remember being a little wonky... so... I said three... let’s see...what would be the case here? I don’t know if all of these three forces are the same... that’s the issue. I don't know if this is applying to each one in particular... like if it's 12,000 here, here, and here... or if it's split evenly? I hope it's being split evenly... that’s what I’m assuming.”} P-F03 Think-aloud
\end{quote}

Although participants were aware that the system might interpret the load assignments differently, only a few were actually able to figure out and correct the mistake.

\subsection{Learning strategies among successful\\ designers \rev{(RQ1b)}}

\rev{
Here we present findings related to \textit{how designers overcame the previously reported learning challenges} \textbf{(RQ1b)}.
We observed that all designers (after watching the introduction tutorial video) tried to learn to work with the tools through an iterative trial-and-error process.
We also observed that participants sporadically consulted different support resources, including software tooltips and help files, and external resources like video tutorials, online user forums, and in some cases asking colleagues for help.
Designers sought help from these support resources primarily after encountering interface or usability issues, which they often resolved. However, despite the available support resources, most designers struggled to learn to co-create with the tools effectively. Nonetheless, some designers employed successful strategies that helped them in learning to work better with the AI systems:
}

\begin{table*}[t]
\caption{Overview of Study 2: Learning with a Peer Guide participants. Six additional participants (3xFusion360, 3xSimuLearn) were paired with peer guides who had worked successfully with the AI design tools in Study 1.
}
\label{tab:participants_guided}
\begin{tabular}{llccllccc}
\toprule
\textbf{ID}    & \textbf{Group} & \begin{tabular}[c]{@{}l@{}}\textbf{Age} \\ \textbf{Years}\end{tabular} & \textbf{Gender} & \textbf{Domain} & \textbf{Occupation}                 & \begin{tabular}[c]{@{}c@{}}\textbf{CAD Exp.} \\ \textbf{Years}\end{tabular} & \begin{tabular}[c]{@{}c@{}}\textbf{Ind. Exp.} \\ \textbf{Years}\end{tabular} & \begin{tabular}[c]{@{}l@{}}\textbf{Guided by} \\ \textbf{Participant}\end{tabular}\\
\midrule
P-F07 & Fusion360 & 27 & M & Mechanical Engineering & Student / PhD & \textgreater 5 & 1 – 2 & P-F05 \\
P-F08 & Fusion360 & 27 & M & Mechanical Engineering & Student / PhD & \textgreater 5 & 1 – 2 & P-F05 \\
P-F13 & Fusion360 & 19 & M & Mechanical Engineering & Student / BA  & 2 – 4          & 0     & P-F11 \\
P-S05 & SimuLearn & 26 & F & Architecture           & Student / MA  & \textgreater 5 & 2 – 5 & P-S04 \\
P-S08 & SimuLearn & 20 & F & Architecture           & Student / BA  & 2 – 4          & 0     & P-S04 \\
P-S09 & SimuLearn & 23 & F & Architecture           & Student / BA  & 2 – 4          & 2 – 5 & P-S04 \\                                                

\bottomrule
\end{tabular}
\end{table*}

\subsubsection{\textbf{Systematically exploring AI's limitations and capabilities}}

We observed that, early on in their interactions, two designers (P-F11, P-S10) deliberately and systematically experimented with the AI tools to develop a better intuition of the AI's behavior, capabilities, and limitations. These designers conducted tests to understand what effects different parameter values would have on the final result and documented the value-result correspondence to create a mental model. For example, PS-10 realized that their initial design sketches were not feasible with SimuLearn and the shape-changing materials, thus decided to systematically test different extreme grid shapes to hone their mental model of the AI's behavior:

\begin{quote}
    \textit{"Even though I tried sketching some stuff, I think it just didn't work. So I thought it's better if I just go into the tool and see if I will be able to do this. I tried stuff like folding one corner upwards and one corner downwards or stuff like that. I took lots of screenshots and those really helped me to understand like 'if I do this, then it's gonna behave like that' so I think initially it was a lot of trying to form a mental model and like what's the capability of this tool."} P-S10 Interview
\end{quote}

\subsubsection{\textbf{Sketching, explaining and reflecting on design issues} }

Another strategy that helped designers overcome flawed outcomes was to actively abstract and explain the problem. In the think-aloud sessions, we observed when facing similar fundamental challenges like over-constraining the engine bracket (such that the loads had no effect), some designers were able to understand and overcome the issue by sketching out the bracket's free-body diagram and explaining the acting forces and constraints to themselves. We observed similar strategies in SimuLearn, where participants understood the significant influence of gravity during the transformation process by explaining the process to themselves: 

\begin{quote}
    \textit{"Whoa, I was not expecting that at all... uh... is that just because of gravity? And there's something crazy going on here... there's a lot of gravity... oh is it because I made the thing so big? Yikes, that is not at all what I expected."} P-S11 Think-aloud
\end{quote}

\section{Study 2: Learning with a\\ peer guide \rev{(RQ2)}}
\label{sec:guided-sessions}

\rev{To gain insights into \textit{how designers can be better supported in learning to co-create with computational AI tools} \textbf{(RQ2)}, we conducted additional think-aloud sessions where designers were paired with experienced peers to support them during the task (i.e., guided sessions). The guide had prior experience using the AI tools and provided support as needed to help participants to more effectively co-create with AI-driven features. Motivated by human-human collaboration, we aimed to derive insights into when and how to effectively support users in learning to co-create with AI by observing the support strategies, pedagogical moves, and communication patterns of the human guide.}

\subsection{Method}
We recruited six additional participants (Table \ref{tab:participants_guided}) following the same criteria as described in Section \ref{sec:participants}. \rev{The guides were recruited from the pool of participants} who had completed Study 1 and demonstrated a thorough understanding of the domain, tool, and task (P-F05, P-F11, P-S04). We asked these guides to support the designers in learning to co-create with the tool. No script was provided to the guides because we intended to find possible support strategies from their natural interactions.

All guided sessions followed the same procedure as that of the unguided sessions, except that the homework sessions were limited to 50 minutes and the guides were present to help the designers. The designers and the guides communicated with each other through audio, screen sharing, and screen annotations. A researcher quietly observed and took notes. After the design session, we conducted separate 15-minute semi-structured interviews with the guide and designer. All sessions and interviews were recorded (video and audio) and automatically transcribed. 

We conducted a reflexive thematic analysis to identify situations and themes on how peer guides supported designers to overcome challenges in learning to work with AI tools. \rev{We specifically focused on aspects of collaboration and knowledge transfer such as communication, joint information processing, and coordination of actions \cite{meier_rating_2007}.} We coded the video, think-aloud transcripts, and interview transcripts then generated themes \rev{with a focus on communication and learning} by analyzing the guide's actions and support strategies that helped designers overcome challenges previously observed in Study 1.

\subsection{Results – Guided Sessions}

The guides supported the designer in using the tools, understanding the AI's behaviors, capabilities, and limitations, and sometimes suggesting and discussing alternative design goals or strategies. Guides primarily reacted to designers' verbalization and actions when they asked for help, expressed uncertainty, or when guides observed common mistakes. To get an impression of conversational dynamics, please see the video figure in the supplementary material.
We portray five of the most common peer support strategies that helped designers to learn better to co-create with AI systems: 

\subsubsection{\textbf{Guide providing step-by-step walk-through instructions}}

Guides often provided designers with step-by-step instructions for setting specific parameters. Such instructions were provided in response to designers' actions, such as showing confusion or struggle, but sometimes designers also specifically requested such assistance.

\subsubsection{ \textbf{Guide reacting to designers' expressions of uncertainty} }

We observed that guides were especially sensitive to moments when designers signaled uncertainty or when designers verbalized knowledge gaps with hedging expressions such as \textit{"I don't know"}, \textit{"maybe"}, or \textit{"I assume."} In these situations, the guides often intervened and offered support or suggested alternative design strategies. Here, the guide suggested creating an obstacle geometry in Fusion360 to prevent material build-up at the bottom of the part in response to the designer who was wondering how to keep the part within the specified dimensions.

\begin{quote}
    DESIGNER: \textit{"...but because the bolt is not here, I don't know where it would be..."} \newline 
    GUIDE: \textit{"I mean maybe on the bottom face, right? That's what's resting on the body of the ship effectively."} \newline 
    DESIGNER: \textit{Yeah, let's go ahead and choose that bottom face."} \newline
    (Designer P-F13, Guide P-F11)
\end{quote}

\newpage
\subsubsection{\textbf{Guide prompting designer reflection on generated designs}}
We often observed situations where the guides prompted feedback from designers each time SimuLearn's simulation or Fusion360's solver had finished. 
In those moments, peer guides often asked the designer \textit{“Is that what you envisioned?”} or \textit{“Is that what you wanted?”} or even \textit{“Yeah, there you go! Is that how you want it to bend?”}. These prompts triggered designers to reflect on the generated designs, which helped the guides better understand how to provide support. 

\subsubsection{\textbf{Guide suggesting alternative means and goals}} 

Beyond supporting tool operation and technical troubleshooting, the guides frequently suggested alternative means and goals to the designer, as in this dialogue:

\begin{quote}
    DESIGNER: \textit{"It's going for the green part, but it's not able to figure out like a perfect way to get there without like touching this obstacle geometry."} \newline
    GUIDE: \textit{"Right, well I mean just make your obstacle geometry really long, say, all the way back past the connection to the ship, and then you're saying 'no material is allowed to go here' and that would make sure that you can always get a bolt in. See what I'm saying?"} \newline
    DESIGNER: \textit{"Yeah, I do"} \newline
    (Designer P-F13, Guide P-F11)
\end{quote}

Here the guide suggested an alternative way to achieve the designer's goal by enlarging the existing obstacle geometry. This strategy helped designers to better communicate design goals to the AI system and develop an intuition for harnessing the AI's capabilities. 

\subsubsection{\textbf{Guide and designer making use of screen annotations and mouse gesturing to discuss design strategy}} 

An essential part of building understanding between the designer and the peer guide was through nonverbal communication, such as screen annotations, sketches, or mouse gestures. Guides frequently used the screen annotation feature built into Zoom to highlight elements they spoke about by circling or drawing arrows.
We also observed that all designers naturally used the mouse cursor to emphasize design features through circling or pointing gestures when explaining something to the guide. Both behaviors are illustrated in this situation where the guide and the designer discussed a strategy to achieve a specific bottle holder shape:

\begin{quote}
    GUIDE: \textit{"So in this case, maybe I suggest that you move these two points in particular more towards the center."}\newline 
    \texttt{[Guide draws arrows from points towards the center of the grid] }\newline
    DESIGNER: \textit{"And this one seems to have dropped downwards. Even this point here... The beam seems to be going downward."}\newline 
    \texttt{[Designer points with mouse at different beams]} \newline
    GUIDE: \textit{"oh this beam right here?" }
    \texttt{[Guide draws an arrow pointing at beam]}\newline
    DESIGNER: \textit{"Yeah."} \newline
    (Designer: P-S08, Guide: P-S04)
\end{quote}

Many designers and guides also annotated generated designs and sketched to clarify or illustrate their ideas. In summary, different forms of nonverbal communication helped designers and peer guides develop better shared mental models of the task and collectively overcome design issues.

\section{Designers’ needs and expectations\\ for co-creating with AI-based\\ design tools \rev{(RQ3)}}

Based on our observations and interviews from both the unguided and guided sessions, we highlight four themes that capture the \textit{needs and expectations that designers expressed around co-creating with AI-based design tools} \rev{\textbf{(RQ3)}}. 

\subsection{Designers expect the AI system to have\\ more contextual awareness about the\\ design problem at hand}

In both systems, participants missed the kind of contextual awareness that a human collaborator might have about a design task, such as a part's function or how the part interfaces with other elements in the environment.
Such a lack of contextual awareness was one of the main reasons people thought working with the tool was not collaborative. 
This lack of context also meant that the tool could not support the designers more proactively like a human partner, as described by this participant: 
\begin{quote}
    \textit{"Certainly it would have saved me some time if at the beginning the software would have said 'oh I see that these are your connection points. Can you actually get a bolt in there?’ […] Things like that would have felt really much more collaborative and helpful."} P-F11 Interview
\end{quote}

Others expected the tool to offer more intelligent manufacturing and material suggestions or help them anticipate real-world design issues, as this participant expressed: 
\begin{quote}
    \textit{"I would like to see a design tool that would show me simulations of the water bottle in action, like 'oh is there enough friction’ or 'will it actually stay in place while a cyclist is on the bike' and then provide suggestions of how to alleviate those problems."} P-S04 Interview
\end{quote}

\subsection{Designers desire a more conversational form of interaction with the tool}

Most participants complained about the lack of reciprocal interaction between them and the tool. 
Participants compared designing with the AI systems to \textit{"programming"} or  \textit{"working with a skilled teammate who is not listening to you."} 
Participants wished for a more conversational interaction with the tools, as desired by this designer: 
\begin{quote}
    \textit{“More like a tool that I can have a conversation with while I'm always sure that everything that I'm making is fulfilling the expectation of the piece and the loads and materials and everything.”} P-F04 Interview
\end{quote}

\subsection{Designers require support in thinking\\ through design problems}

Across all guided sessions, designers appreciated that the peer guide helped them learn to operate the tool but also to \textit{think through design problems}---a feature they would eventually also expect from a co-creative AI tool. As summarized by this designer who reflects on working with their peer guide:

\begin{quote}
    \textit{"I think working together with the peer guide was actually really helpful because he had a lot of insight. I felt like a couple of things that I didn't really think about just from a fundamental engineering standpoint of how the thing can actually be made. I think that was really beneficial."}  P-F13 Interview
\end{quote}

\subsection{Designers expect an AI tool to provide \\project-relevant work examples }

Designers frequently suggested that a co-creative tool should suggest similar task- and project-related examples created by other users. 
Such examples could help designers learn the system's capabilities and provide creative inspiration, as suggested by this SimuLearn user:
\begin{quote}
    \textit{"Let's say I'm modeling some form and in real-time, the tool is searching some sort of database to show me some possibilities that other people have previously done, just by the similarity of the shape, and then I'm like ‘yeah this shape can go there and go there.' So that's actually a creative input that can help---a bit more like you're designing with someone. It doesn't necessarily have to generate, it can pull up from other people and tell you 'here is how some other designers work with this.'”} P-S10 Interview
\end{quote}

\begin{table*}[h]
  \centering
    \caption{Overview of design opportunities and example applications in relation to group cognition and team learning.}
    \label{tab:opportunities}
  \includegraphics[width=0.88\textwidth]{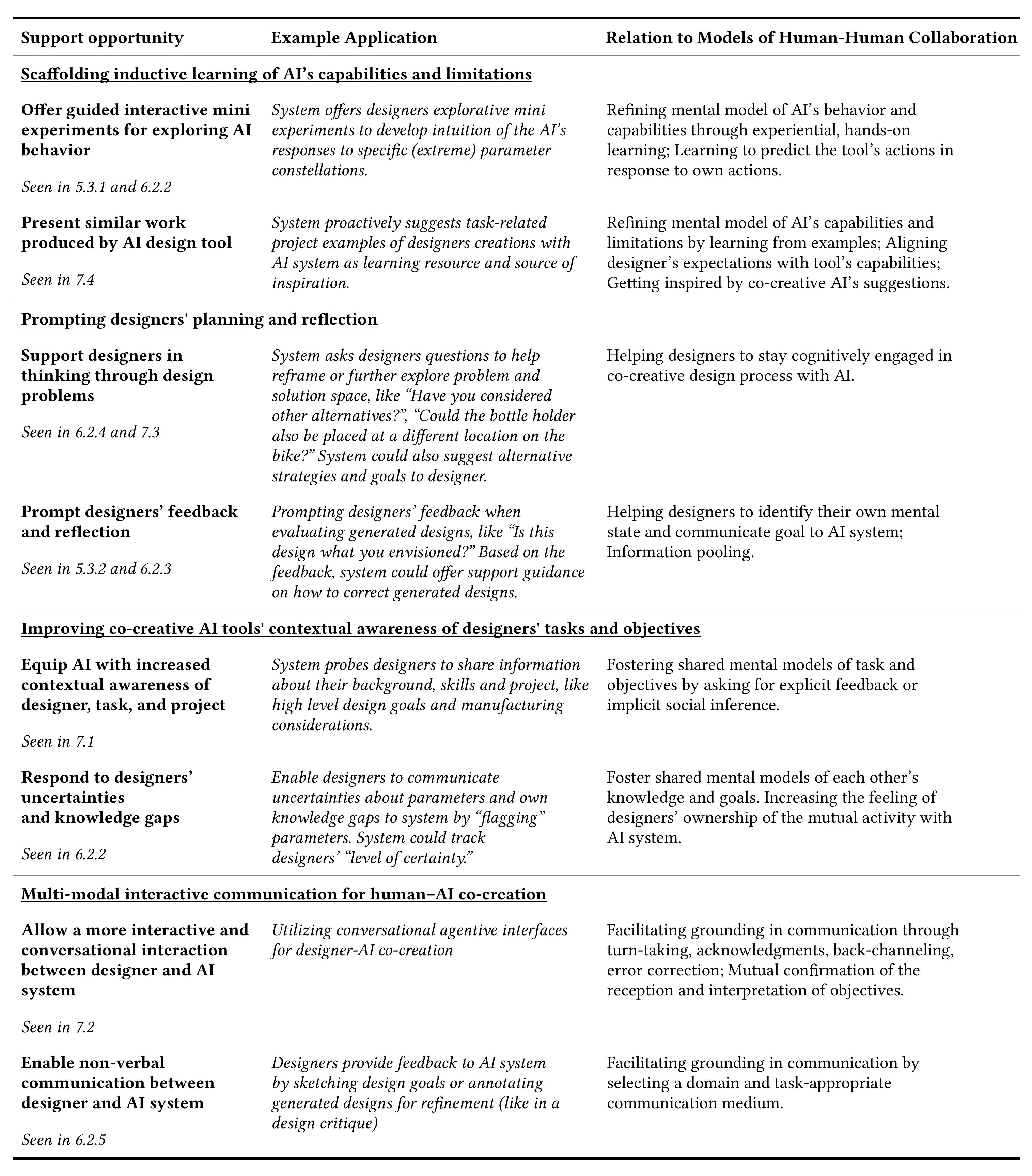}
\end{table*}

\section{Discussion}
Co-creative AI design tools have the potential to amplify the abilities of engineering and industrial designers.
However, we found that designers face major challenges in (learning to) effectively co-create  
\rev{through understanding and adjusting system outputs, communicating design goals, and working "collaboratively" with the AI.
Designers who overcame challenges did so by systematically exploring AI's limitations and capabilities, and by explaining and reflecting on their design issues. The observed support strategies of peer guides comprised step-by-step walk-through instructions, reacting to designers' expressions of uncertainty, prompting designer reflection on generated designs, suggesting alternative means and goals, and making use of screen annotations and mouse gestures to discuss design strategies. Overall, designers expected the AI system to have more contextual awareness about the design problem at hand, desired a more conversational form of interaction with the tool, asked for more support in thinking through design problems, and expected project-relevant work examples from the tool. }

We discuss our findings in the context of prior research studying how users learn to work with complex software tools and relate our findings to concepts and empirical findings from research on \rev{human collaboration}.
We highlight design opportunities (Table \ref{tab:opportunities}) to better support designer-AI co-creation by scaffolding designers in actively exploring the boundaries of AI capabilities and limitations, prompting designers to actively reflect on design problems and observed AI behaviors, enhancing AI systems' contextual awareness of designers' tasks and objectives, and supporting more conversational forms of multi-modal communication between designers and AI systems. 

\subsection{\rev{New learning challenges for\\ human--AI co-creation}}

\rev{Based on our findings, we believe that the challenges designers faced when learning to co-create with the tools go beyond learning the tools' interfaces.
Firstly,} all participants were experienced designers familiar with the CAD software's interface and had watched a step-by-step instructional video demonstrating the operation of the AI features.
Furthermore, designers used different support resources, including tooltips, help files, video tutorials, online user forums, and asking colleagues for help \rev{to overcome interface issues they faced.}
However, even with the help of these resources, most designers could not produce feasible and satisfying outcomes with AI assistance. While this result may have been due, in part, to the limitations of the AI tools themselves, it is clear that this was not the full story. Indeed, a few of our participants were able to overcome challenges and learned to co-create effectively with the AI features. Hence we believe that the challenges participants faced when working with the AI systems (such as communicating design goals or understanding AI outputs) were partly \textit{new learning challenges} due to the tool's more active role in the design process. 
\rev{These learning challenges go beyond learning the tool's interface. For example, a major challenge we identified relates to how designers specify all required parameters upfront instead of modeling and testing a part step-by-step. This workflow requires designers to think through the design problem in advance, which is challenging and different from the usual iterative design process. }
Our findings suggest that these new learning challenges require new support strategies for example, by prompting designers' reflection in response to expressions of uncertainty or suggesting alternative design goals that align with the AI's capabilities.  

\subsection{Toward \rev{models of human collaboration}\\ as lenses for studying and designing\\ co-creative systems}
\label{Toward group cognition and team learning as lenses for studying and designing co-creative systems}

Participants had trouble learning to predict how the AI might behave in response to the specified parameters.
They struggled to make sense of the AI system’s reasoning and struggled to correct unwanted design issues.
Prior literature on group cognition suggests that to achieve effective collaboration group members should be able to interpret each other's reasoning and predict roughly how their partner might behave in response to their own actions \cite{zhou_group_2018}. 
Similarly, from a team learning perspective, our findings suggest that designers who systematically explored the AI’s limitations and capabilities early on were better at predicting the tool’s actions in response to their own and produced more satisfactory results.
This result is in line with studies on human-AI collaboration in decision-making, suggesting that users learn to better predict the machine's behavior through inductive mechanisms (i.e., via concrete examples and hands-on testing) than via general, declarative information about internal processes \cite{chandrasekaran_it_2017}. \rev{While explainable AI research focuses primarily on directly communicating information about the AI system to the user, recent research has suggested that more engaging and longer forms of learning and deliberate practice might improve human-AI collaboration \cite{kawakami_towards_2022}. 
However, in addition to supporting honing the user's mental model of the AI's capabilities and limitations, it is equally important for the AI system to have an understanding of the user's capabilities, limitations, and task context to enable more effective human-AI collaboration.  
Hence, this would require the AI system to have better contextual awareness of the user and the current task at hand. We further discuss the resulting design opportunities in section 8.3.3. } 

Most designers felt the tools were uncollaborative and had more control over the design process than they would have preferred.
As a consequence, they accepted imperfect results, developed improvised workarounds, or avoided AI assistance altogether. 
Previous studies on group cognition and team learning suggest that group members feel more ownership of the mutual activity when the group learns to coordinate cognitive capabilities among participants---united by their interpretations of each other’s mental states \cite{mohammed_cognitive_2001}.
Groups can reach this level of collaboration by an active process of communication, joint information processing, and coordination of actions \cite{meier_rating_2007}.

Our findings from the guided sessions show that the guides’ active communicative support strategies, such as reacting to designers’ uncertainties and providing step-by-step instructions, helped designers learn to work confidently with the AI.
Furthermore, when peer guides prompted designers' feedback and reflection on generated designs, the designers were required to articulate their intentions explicitly. As a result, both the guides and designers were able to discuss and better coordinate further actions to improve the outcome.

\subsection{Design Opportunities and Future Work}

\rev{The following section highlights several design opportunities we identified in our findings to support designers in learning to co-create with computational AI tools.}

\subsubsection{\textbf{Scaffolding inductive learning of AI’s capabilities and limitations} }


As discussed, some designers learned the AI system's capabilities and limitations by testing and documenting the effectual correspondence between various parameters and the generated result (see 5.3.1). In the guided sessions, peer guides also frequently provided step-by-step instructions for setting up parameters or walked designers through the sequence of steps (see 6.2.2). These inductive learning strategies helped designers to better predict the AI’s behavior and understand its capabilities.
\rev{Previous work on novice-AI music co-creation has also found that users systematically tested AI limitations to hone their mental model of the system's behavior \cite{louie_novice-ai_2020}. Going further, to better support users in this learning activity,} future co-creative systems may offer designers a set of hands-on mini-guided ‘experiments’ to better understand the system’s responses to specific (extreme) parameter inputs \cite{koedinger_knowledge-learning-instruction_2012}. 
Systems may also offer designers opportunities to view sets of examples of input-output pairs to help designers develop useful mental models of an AI tool's generative capabilities and limitations (cf. \cite{mozannar_teaching_2021}).
A co-creative tool may also proactively recommend similar tasks and project-related examples created by other human-AI teams (see 7.4) to help designers learn the system’s capabilities and provide creative inspiration.

\subsubsection{\textbf{Prompting designers' planning and reflection }}

Participants who were more successful at co-creating with the AI tools did so by abstracting and explaining their problems---either to themselves during think-aloud sessions or to the peer guides (see 5.3.2 and 6.2.3). Literature from the learning sciences shows that self-explanation positively affects understanding and problem-solving \cite{vanlehn_model_1992}.
In addition, participants from the guided sessions appreciated the guides' prompt for reflection on AI behaviors and their suggestions for alternative design goals or strategies (see 6.2.4 and 7.3). Such actions helped designers think through the tasks and plan actions with AI tools. Conversely, designers who did not reflect on the design problems were unable to learn to understand the system's behavior, capabilities, and limitations well enough and failed to produce satisfactory outcomes. 
\rev{This observation was more prevalent in Fusion360 than in SimuLearn since Fusion360's long simulation time doesn't support the kind of rapid interactive adjustments as SimuLearn did, thus requiring the user to strategize their actions in advance. Hence, supporting users in thinking through the design problem for specifying parameters ahead would be especially beneficial for AI tools with longer processing time.}

One additional explanation for \rev{why participants failed to produce satisfactory outcomes} might be that AI systems can lead designers to over-rely on their support, creating an \textit{``illusion of success''} that reduces their effort in solving the design problem \cite{zhang_cautionary_2021}, something we saw when designers accepted results even when they appeared unfeasible (see Section 5.2.1).
To compensate for this tendency, a co-creative system may help designers reframe the problem or further explore the solution space by suggesting alternative goals or asking \textit{generative design questions} \cite{eris_asking_2003} like \textit{``What could other alternatives look like?''.}
Moreover, actively reflecting on the design process is an essential part of professional design practices \cite{schon_reflective_1983}. 
Much like the guides, a co-creative system could prompt feedback and active reflection on observed AI behaviors or generated designs by asking deep reasoning questions about the results, such as \textit{``Is this generated design what you envisioned?''} (cf. \cite{chase_how_2019}). 
Based on the feedback, the system may then offer support and help to coordinate further actions. This strategy would be complementary to the inductive learning support described in the previous section (8.3.1).

\subsubsection{\textbf{Improving co-creative AI tools' contextual awareness of designers' tasks and objectives}}

Our results show that designers felt a lack of the tool’s awareness about the design context and therefore missed the kind of proactive support a human partner might provide (see 7.1). 
While building contextual awareness into AI systems has long been a tradition in HCI research, it also presents many technical challenges. However, in the context of co-creative design tools, promising directions are being explored. For example, the system could derive its user model through explicit and implicit mechanisms to develop a shared mental model of the context by asking the user for information about the specific design task (i.e., parsing a written design brief for context and goals \cite{tan_text2scene_2019}) or infer design goals from user behaviors \cite{law_design_2020}. 

From a group cognition and team learning perspective, contextual awareness also includes an understanding of the other team member’s existing or missing knowledge about the design task. Based on our findings (see 6.2.2), a co-creative tool might learn about the designers’ knowledge by responding to verbalized knowledge gaps such as \textit{``I don’t know''} or responding to hedging expressions such as \textit{``maybe''} or \textit{``I assume.''} This observation is in line with research showing student learning is positively affected by human tutors' responses to their expressions of uncertainty \cite{forbes-riley_adapting_2009}. Based on this phenomenon, literature has explored how intelligent tutoring systems can detect and respond to student hedging \cite{forbes-riley_adapting_2009,pon-barry_responding_2006}.
A co-creative system may also allow designers to communicate uncertainties about parameters and their own knowledge gaps.
For example, designers may flag an ``\texttt{???}'' checkbox next to a parameter’s input field to signal uncertainty. The AI system could then track designers' ``level of certainty'' for each parameter and provide reactive help or re-surface those parameters later in the design process to identify possible reasons for unexpected outcomes.

\subsubsection{\textbf{Multi-modal interactive communication for human–AI co-creation}}

Our findings show that designers felt communicating with the AI was like giving instructions without receiving much feedback (see 7.2). Results from the guided sessions show that peer guides used conversational strategies like confirming the reception of each other's goals or asking for clarifications when they were unsure about the other's intentions. As conversation is widely seen as a vital mode of designing \cite{schon_reflective_1983} and empirical work suggests that much of design work lies within conversations between collaborators and stakeholders \cite{lawson_computers_1997}, co-creative systems should consider using more back-and-forth conversation as an interaction interface. 
Furthermore, studies on team learning show that the forming of effective, shared mental models is strengthened through an active process of iterative negotiation between team members, involving “constructive” forms of conflict, argumentation, and resolution \cite{head_effective_2003,jeong_knowledge_2007,van_den_bossche_team_2011}. Such a strategy may also prove useful for negotiating design goals.

Our findings from the guided sessions also suggest that non-verbal communication may support design partners in developing better shared mental models of design goals (see 6.2.5). 
Designers and guides discussed goals and strategies by pointing at features with the mouse cursor or sketching with the screen annotation feature, similar to a \textit{'spatial-action language'} described by Schön \cite{schon_reflective_1983} which explains gesturing and drawing along with verbal expressions as typical forms of communication in traditional design critique sessions. 
Such non-verbal interaction is still an underutilized medium in human-AI co-creation.
Allowing designers to use sketching, annotation, or gesturing atop generated results may help them communicate design goals to an AI system.

\subsection{\rev{Human--AI co-creation beyond\\ manufacturing}}
\rev{
Although we identify opportunities to support learning to co-create with AI systems in the context of manufacturing, many of our findings could also apply to other human-AI co-creation domains such as image, music, or text generation. 
Given recent advancements in generative image AI models (such as DALL-E \cite{ramesh_zero-shot_2021} or Stable Diffusion \cite{rombach_high-resolution_2022}) with fast release cycles of new tools and capabilities, supporting creative professionals in learning to effectively co-create with such tools might become increasingly important. Furthermore, many prompt-based AI models like DALL-E expect users to express their goals through text prompts, which is an unfamiliar modality for most of today's visual designers. 
Consequently, our findings suggest that AI tool users across many domains could be supported in learning to better co-create with AI systems by scaffolding inductive learning of the AI’s capabilities, prompting users’ planning and reflection, improving the tools’ contextual awareness of tasks and objectives, and facilitating multi-modal interactive communication between tool and user. 
Future work might further explore interfaces for supporting learning to co-create with computational AI tools in other domains beyond manufacturing. }

\subsection{Limitations of the study}

We highlight three limitations of this study: first, our participants only represent a subset of engineering, architectural, and industrial designers. Although all participants had relevant training in their design fields and worked with 3D CAD software, most had minimal industry experience. We also included three professionals with substantial industry experience to compensate for this imbalance, however, even these participants struggled. Further, the self-selected participants in our study were presumably interested and open  to the idea of co-creating with an AI system. Thus, some of our findings may be reflective of this openness to co-creative work. 
Second, both AI tools are still early in development, and we noticed user experience issues that could benefit from improved UIs. To our best extent, we isolated UI issues from more fundamental challenges in learning to co-create with AI systems. Although the tools are relatively new, we believe our findings provide value in exploring new opportunities for co-creative design systems. Finally, despite our effort to make sure the design tasks were realistic, designers knew that they were in a research study and that the designs would not be manufactured. In a professional context, participants might have spent longer learning the tool to produce feasible designs.

\section{Conclusion}

In this paper, we presented an empirical study to understand how engineering and architectural designers learn to work with AI-based manufacturing design tools, documenting their challenges in working with an AI and probing their needs and expectations for co-creating with AI systems. 
We identified several support opportunities with an eye toward learning from effective human-human teams to improve future designer-AI co-creation. 
Overall, we aim to inspire others to explore untapped support opportunities and to work toward future co-creative tools that combine the strength of both human and AI systems to achieve complex designs that neither could achieve alone. 

\begin{acks}
We want to thank all study participants, peer guides, and the research assistants Anita Sun, Linda Xue, and Sophia Timko for supporting this work. 

This material is based upon work supported by the National Science Foundation under Grant No. 2118924 Supporting Designers in Learning to Co-create with AI for Complex Computational Design Tasks. Any opinions, findings, and conclusions or recommendations expressed in this material are those of the author(s) and do not necessarily reflect the views of the National Science Foundation.
\end{acks}

\bibliographystyle{ACM-Reference-Format}


\appendix

\newpage
\onecolumn
\section{Additional Materials}

\begin{figure}[h]
  \includegraphics[width=\textwidth]{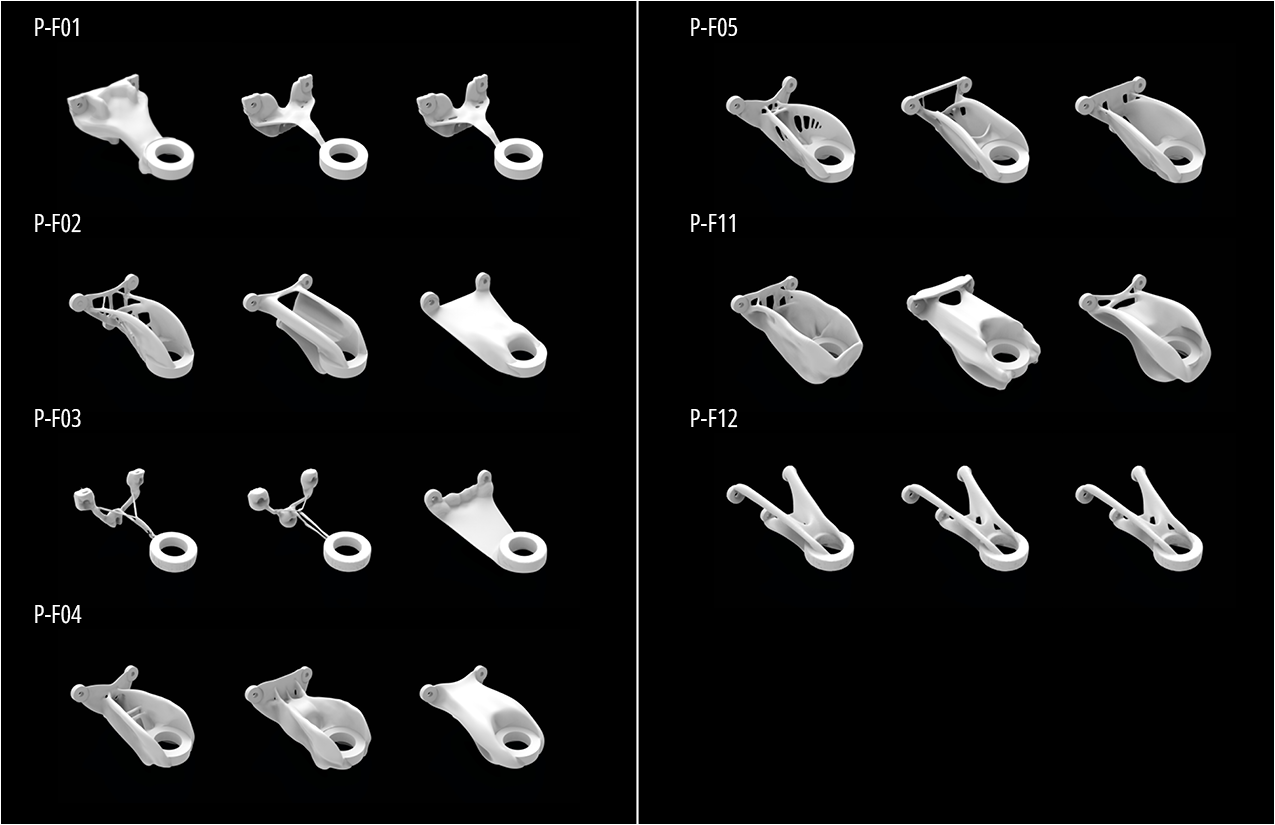}
  \caption{Overview of all engine mounting brackets created in the Fusion360 task. Every participant submitted three design options.
Since the AI system generated the three options based on the same parameter values, each trio either met or missed the same criteria.}
  \Description{Image showing renderings of all 21 engine brackets that were created by participants. Arranged as seven triples on black background with participant codes on the left. }
\end{figure}

\twocolumn
\begin{table*}
  \caption{Interview protocol with questions of the semi-structured post-task interview.}
  \includegraphics[width=0.82\linewidth]{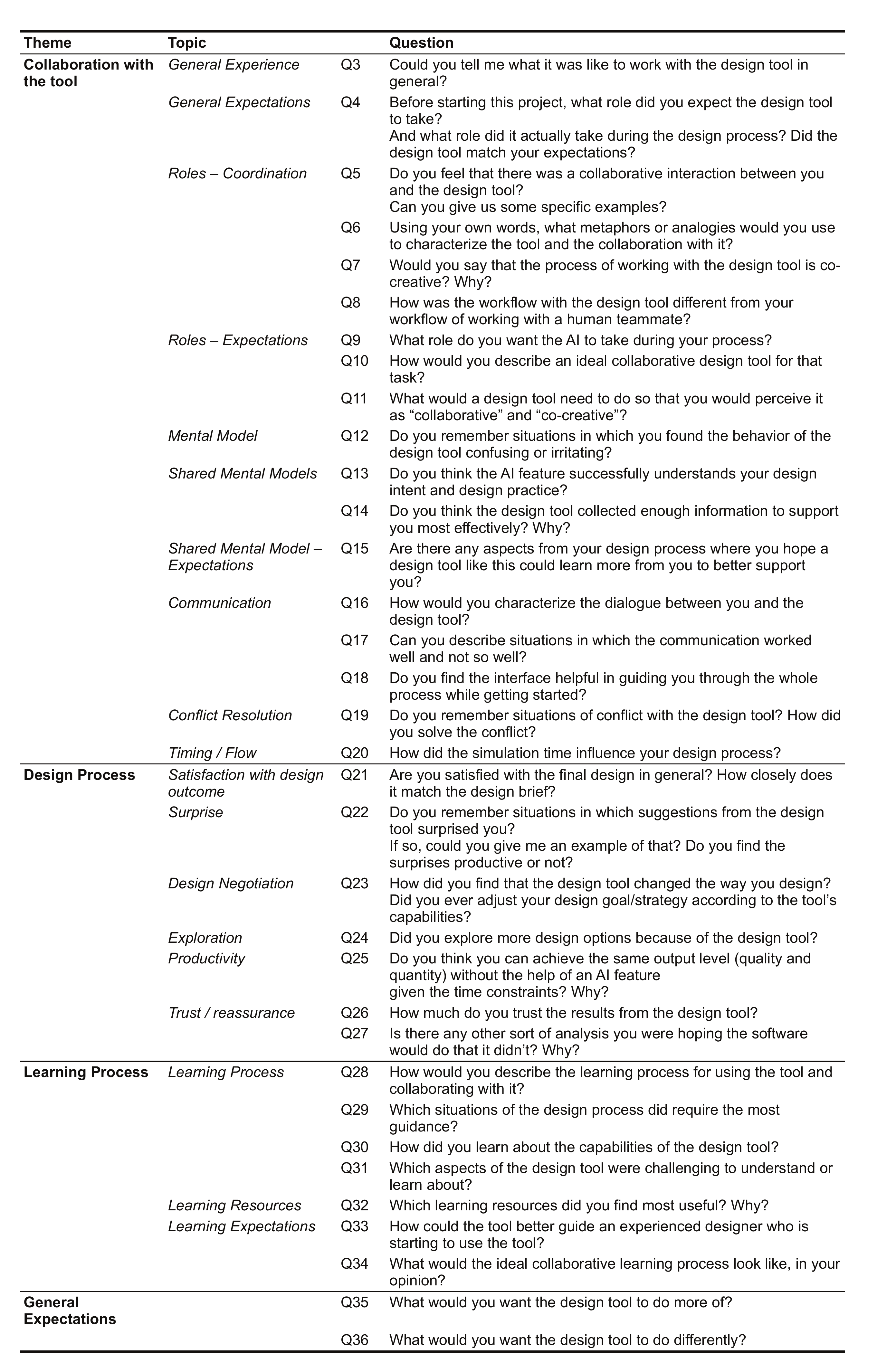}
  \Description{Interview protocol with 36 questions of the semi-structured post-task interview clustered in themes “Collaboration with the tool,” “Design Process,” and “Learning Process.”}
\end{table*}

\end{document}